\documentclass[]{aa}  

\usepackage{graphicx}
\usepackage{txfonts}
\usepackage{colortbl} 
\usepackage[dvipsnames]{xcolor}
\usepackage{epstopdf}
\usepackage[rightcaption]{sidecap}

\begin{document} 

\title{Angular momentum and lithium transport from main sequence to sub-giant and red giant low-mass stars}

   \authorrunning{T. Dumont}
   
   \titlerunning{Angular momentum and lithium transport from main sequence to sub-giant and red giant low-mass stars}
   

   \author{T. Dumont
          \inst{1,2} \fnmsep\thanks{e-mail: thibaut.dumont@iphc.cnrs.fr}
          }

   \institute{University of Strasbourg, CNRS, IPHC UMR 7178, F-67000 Strasbourg, France
         \and Department of Astronomy, University of Geneva, Chemin Pegasi
   51, 1290 Versoix, Switzerland
             }

   \date{(Received; Revised; Accepted)}

 
  \abstract
   {Asteroseismology provides a unique opportunity to probe the interiors of evolved stars and constrain their internal rotation. The correct reproduction of the core rotation evolution has not yet been achieved, although it is key to understanding the internal processes involved in low-mass stars.} 
   {We explore the efficiency required to reproduce the general behaviour of the transport of angular momentum along the evolution in view of asteroseismic constraints from giant low-mass stars. We analyse the consequences and predictions for lithium and beryllium surface abundances from the main sequence to red giant phase.  
   }
   {We computed a series of models, which included atomic diffusion, rotation-induced mixing, magnetic braking, and additional processes tailored for main sequence low-mass stars. We extended these models to more evolved phases and investigated an updated angular momentum transport by including a time-dependent extra viscosity related to the azimuthal magneto-rotational instability. We compared our predictions to the asteroseismic measurements of the core and surface rotation of a sample of sub-giant and red giant stars. We compared the model predictions for the lithium and beryllium surface evolution with the available observations.
   }
   {We confirm that a time-dependent additional viscosity $\nu_{\rm add}(t)$ is required to reproduce the general behaviour of the core rotation rate along successive stellar evolutionary phases given the dependence on the differential rotation and the mass. We show that it results in stronger lithium and beryllium depletions for low-mass stars over evolution. We confirm that predicted lithium abundances at the red giant bump by classical models, commonly used as references, cannot reproduce the lithium depletion along the main sequence and evolved phases of stellar evolution. We show that the observed amount of lithium of stars less massive than 1M$_{\odot}$ leads to a discrepancy between model predictions and observations at the red giant bump. 
   }
   {We show that a semi-parametric model can reproduce the rotational behaviour along the first phases of evolution well, with the exception of the sharp transition observed during the sub-giant phase. This suggests that two distinct transport processes may be involved. The processes required to transport chemicals during the main sequence phase and angular momentum until the red giant phase impact the lithium depletion all along the evolutionary duration. A good prediction of the lithium abundance at young phases places strong constraints on the predicted one at more evolved phases. It also highlights discrepancies between models and observations for the lowest mass stars and impacts the threshold that defines  lithium-rich giant stars, showing that classical models tend to overestimate this threshold.
   }


   \maketitle
   
%

\section{Introduction}
\label{section:introduction}
The description and understanding of the internal transport processes along evolution of low-mass stars is at the heart of vast significant progress in stellar astrophysics. Low-mass stars have been observed to show an important rotational and chemical evolution involving different transport processes that still have to be clearly identified and characterised. The exploration of the transport of angular momentum and of the transport of chemicals along evolution is a key in order to constrain these processes, thanks to the numerous observations that have been obtained for low-mass stars along the pre-main sequence (PMS), main sequence (MS), sub-giant branch (SGB), and red giant branch (RGB) phases. In particular, the internal rotation has been inferred by helio- and asteroseismology \citep[e.g.][]{2012A&A...548A..10M,2012ApJ...756...19D,2014A&A...564A..27D,2015A&A...580A..96D,2020A&A...641A.117D,2013A&A...555A..54C,2013A&A...549A..74M,2015MNRAS.452.2654B,2018A&A...616A..24G,2019LRSP...16....4G}. Observations show that a strong coupling takes place during the MS from an important differential rotation in the young ages to an almost uniform rotation when reaching the solar age and the late steps of MS \citep[e.g.][]{2015MNRAS.452.2654B}. Efficient transport of angular momentum is then expected to take place along MS until the beginning of the SGB phase. When reaching the SGB phase and beyond, along the RGB, the stellar structure is changing, the envelope is expanding, and the core is contracting on itself. This leads to the acceleration of the core rotation and the slowing-down of the surface rotation, driving an increase in the differential rotation. In non-standard stellar evolution code including rotation, the transport of angular momentum usually comes from the combined action of meridional circulation and turbulence shear \citep[e.g.][]{1992A&A...265..115Z}. However, these processes drive a strong differential rotation and are not able to reproduce the almost flat rotational profile of the Sun (MS phase) as well as the moderate core rotation rate observed during SGB and RGB phases thanks to helio- and asteroseismology \citep[][]{2012ApJ...756...19D,2014A&A...564A..27D,2020A&A...641A.117D}. In addition, the transport of angular momentum have been shown to be varying along evolution with an efficient transport during the MS, then a decreasing phase along SGB, and, finally, a new increase during the RGB phase \citep[e.g.][]{2014A&A...564A..27D,2020A&A...641A.117D,2016A&A...589A..23S,2019A&A...626L...1E,2019A&A...621A..66E,2022A&A...663A.180M,2023arXiv230207811M}. These observational constraints suggest the action of an additional unknown transport of angular momentum whose identification remains an open question. \\ The characterisation of the transport of angular momentum along low-mass star evolution is consequently challenging and may potentially involve different processes taking place at different phases with different efficiencies. Since the fact that one (or more) transport processes is lacking has  been demonstrated, there have been several dynamical processes proposed, mainly those related to rotation, such as the internal gravity waves \citep[e.g.][]{1993A&A...279..431S,2005Sci...309.2189C,2005A&A...440..653M,2013A&A...554A..40C,2017A&A...605A..31P}, magnetic fields and instabilities \citep[e.g.][]{2002A&A...381..923S,2005A&A...440L...9E,2010A&A...519A.116E,2019A&A...621A..66E,2019A&A...631L...6E,2022NatAs.tmp..119E,2010ApJ...716.1269D,2014ApJ...796...17F,2019MNRAS.485.3661F}, convective pumping \citep[][]{2015ApJ...808...35K}, transport by mixed-modes \citep[][]{2015A&A...579A..31B}, and see a general review by \citet[][]{2019ARA&A..57...35A}. However, a clear identification of the involved processes is still debated and requires a better description of the angular momentum transport evolution thanks to the observational data.

In parallel, the transport of angular momentum has an impact on the internal mixing of chemicals in stars as was shown, for instance, by \citet[][]{1990ApJS...74..501P,1992A&A...255..191C,1996A&A...312.1000R,2009A&A...501..687D,2020A&A...643A.164S,2022NatAs.tmp..119E}; \citet[][hereafter referred to as Papers I and II]{2021A&A...646A..48D,2021A&A...654A..46D}. The evolution of light elements surface abundances is especially sensitive (directly or indirectly) to the processes involved in angular momentum transport. The evolution of the photospheric lithium (hereafter, Li) has been shown to be sensitive to the rotational evolution, and consequently a good tracer of internal processes (see Paper I and reference therein). Its evolution is well described from the galactic to the stellar evolution scale \citep[e.g.][]{1982A&A...115..357S,2000IAUS..198...61D,Prantzos2012b,2021A&A...649L..10C}, as well as along the different stellar evolution phases \citep[e.g.][]{1976PASP...88..353B,2005A&A...442..615S,2017AJ....153..128C,2019AJ....158..163D,2020A&A...643A.164S,2020A&A...633A..34C,2020ApJ...888...28B,2022ApJ...933...58C}. In addition, the Li depletion was observed to be clearly dependent on stellar masses and metallicities \citep[see for instance][]{2000IAUS..198...61D}. In particular, the so-called Li-dip is observed for F-type stars at about 6'600 K in open clusters like the Hyades and older \citep[e.g.][Paper II]{1986ApJ...302L..49B,1993AJ....106.1080S,1995ApJ...446..203B,2016ApJ...830...49B,2022ApJ...927..118B}, witnessing the action of specific transport processes.

Stellar evolution models accounting only for convection as a mixing process (Classical models) predict Li-depletion only during the PMS when the convective envelope is deep enough that Li is burned in the hot layers of the star. However, it does not predict additional variation of surface Li along the MS and until the star reaches the first dredge-up, in contradiction to observations. It has been shown that only models including rotation-induced processes were able to reproduce the correct Li-depletion over evolution \citep[e.g.][Paper I]{2003A&A...399..603P,2016ApJ...829...32S,2020A&A...633A..34C,2022NatAs.tmp..119E}, impacting predictions for the different stellar evolutionary steps, and pointing the inconsistency of classical models. 

In this work, we explore the evolution of the angular momentum transport efficiency from MS to SGB and RGB phases, and how it impacts the Li (and beryllium) surface abundance evolution. We extend models, that we already explored and validated for the specific case of the MS solar-type stars in a previous study (Paper I), to evolved low-mass stars at different masses. In Sect.~\ref{sec:obs}, we present the observational data that we used to constrain the processes involved in our models. In Sect.~\ref{STAREVOL}, we describe the input physics of the stellar models and the different processes implemented in the evolution code STAREVOL used for this work. In Sect.~\ref{sect:adjust}, we estimate the transport efficiency that is required to reproduce the core rotation rate of a sample of nine stars (eight SGB and one RGB) for which we have access to the rotation asteroseismic analysis. We study the effects and the relevance of a time-dependent additional viscosity related to the azimuthal magneto-rotational instability (AMRI) defined by \citet{2016A&A...589A..23S}, and successfully used by \citet[e.g.][]{2023arXiv230207811M}, and explore our model predictions. We apply these models for a grid (0.8-1.5 $M_{\odot}$, step of 0.1 $M_{\odot}$) at solar metallicity, and explore the Li and beryllium abundance evolution in Sect.~\ref{sect:grilleli}. In Sect.~\ref{sect:CONCLUSION}, we summarise our results, discuss model predictions, and present our conclusions. 
   
\section{Observational data}
\label{sec:obs}

To constrain the evolution of the angular momentum transport, we gathered a sample of SGB and RGB stars for which core rotation rates constraints are available from asteroseismic analysis. We have no access to stars for which both a seismic analysis and Li abundance determination have been done. We then completed our analysis with the compilation of another sample of low-mass stars for which we have access to Li and beryllium (hereafter, Be) abundances determinations from the MS phase to the RGB bump.

\subsection{Internal rotation}
The internal rotation is estimated from an asteroseismic analysis of the gravity-dominated mixed-modes. The gravity g-modes carry the information about the internal layers of the stars and the internal rotation \citep[see for instance:][]{2013A&A...549A..75G}. Consequently, the determined observational constraints concern the average rotation in the g-mode cavity. In order to compare our model predictions with asteroseismic observations, we define the core rotation $\Omega_c$ as:
\begin{equation}
    \overline{\Omega_{\rm c}} = \int_{0}^{r_N} \frac{\Omega \, dm}{m(r_N)}, 
    \label{eq:Omc}
\end{equation}
with $r_N$ as the top radius of the resonant g-mode cavity determined from the Brunt-Väisäilä frequency, N, and $m(r_N)$ as the cumulative mass at the radius, $r_N$. We used a sample of evolved stars for which the internal rotation have been estimated by mean of a seismic analysis. It includes a sample of eight SGB stars from \citet{2014A&A...564A..27D,2020A&A...641A.117D}, one RGB star from \citet{2016ApJ...817...65D}, and a large sample of RGB stars from \citet{2018A&A...616A..24G}. \\ The effective temperature and surface gravity of the eight SGB stars, and of KIC 44487777, are taken from the reference papers and given in Table~\ref{tab:data}. We used the Mikulski Archive for Space Telescopes (MAST database\footnote{http:// archive.stsci.edu/kepler/stellar17/search.php}) for the sample analysed by \citet{2018A&A...616A..24G}.

\begin{table*}[t!]
    \centering
    \caption{Main properties of the evolved stars considered in this work.}
    \begin{tabular}{c|c|c|c|c|c|c}
         Star & Mass - Reference papers & Mass & [Fe/H] & [Fe/H] & log(g) & $T_{\rm eff}$ \\
         & $M_{\odot}$ (Seismic mass / Best model) & $M_{\odot}$ (This work) & & (This work) & K \\
         \hline \hline
         KIC 5955122 - G & $1.11\pm0.07$ / 1.218 & $1.12$ & -0.17$\pm$0.06 & -0.17 & 3.88$\pm$0.02 & 5865$\pm$70 \\
         KIC 8524425 - H & $1.07\pm0.05$ / 1.113 & 1.10 & +0.14$\pm$0.06 & +0.14 & 3.98$\pm$0.03 & 5620$\pm$70 \\
         A & $1.20\pm0.16$ / 1.22 & $1.20$ & +0.25$\pm$0.23 & +0.25 & 3.83$\pm$0.04 & 5248$\pm$130 \\
         B & $1.27\pm0.15$ / 1.27 & $1.20$ & -0.09$\pm$0.06 & -0.09 & 3.77$\pm$0.02 & 5139$\pm$55 \\
         C & $1.11\pm0.16$ / 1.14 & $1.15$ & +0.24$\pm$0.16 & +0.25 & 3.76$\pm$0.04 & 4978$\pm$167 \\
         D & $1.50\pm0.20$ / 1.26 & 1.15 & -0.15$\pm$0.06 & -0.15 & 3.71$\pm$0.03 & 5264$\pm$60 \\
         E & $1.33\pm0.14$ / 1.39 & $1.50$ & +0.41$\pm$0.06 & +0.41 & 3.68$\pm$0.02 & 5115$\pm$60 \\
         F & $1.24\pm0.17$ / 1.07 & $1.10$ & -0.40$\pm$0.08 & -0.40 & 3.60$\pm$0.02 & 5120$\pm$55 \\
         KIC 4448777 & $1.02\pm0.05$ / 1.13 & 1.15 & +0.23$\pm$0.12 & +0.25 & 3.25$\pm$0.03 & 4750$\pm$250 \\
         \hline
    \end{tabular}
    \tablefoot{References. SGB stars: \citet{2014A&A...564A..27D,2020A&A...641A.117D}; RGB KIC 4448777: \citet{2016ApJ...817...65D}}
    \label{tab:data}
\end{table*} 

\subsection{Lithium and beryllium abundances}
\label{sub:Lisuncluster}
In this work, Li surface abundance are coming from spectroscopic data for a sample of two Galactic open clusters including both MS and giant stars. Their names, ages, and metallicities are given in Table~\ref{tab:OCinfos1} along with bibliographical references from which the Li and Be surface abundances were taken. We only take into account non-binary stars with confirmed membership, as mentioned or flagged in the reference papers cited in Table~\ref{tab:OCinfos1}. In order to reproduce the mass-range and evolutionary state of stars with seismic determinations, we completed our sample adding two samples of solar twins by \citet[][]{2019MNRAS.485.4052C} and \citet{2022ApJ...941...21B}\footnote{We exclude star HIP 32673 from our sample as this star may have lived a mass-exchange or a merger with a stellar
companion as described by \citet{2022ApJ...941...21B}} and two samples of evolved stars by \citet[][]{2007AJ....133.2464L} and \citet{2023ApJ...944L...5M}. We kept only those stars with a metallicity between [Fe/H]= +0.10 and [Fe/H] = -0.10. \\ We adopted the meteoritic abundance $\rm A(Li) = 3.31$\footnote{$\rm A(X) = log_{10}(N_X/N_H)+12$, where $\rm N_X$ is the number density of element X.} \citep[][]{1989GeCoA..53..197A} as the original abundance of lithium\footnote{New determinations have been done in the past years: $\rm A(Li) = 3.28\pm0.06$ \citep[][]{2003ApJ...591.1220L}, $\rm A(Li) = 3.26\pm0.05$ \citep{2009ARA&A..47..481A}, however $\rm A(Li) = 3.31$ is still commonly used as an initial reference for open clusters \citep[e.g.][]{2020ApJ...888...28B} and when not focusing on the Sun.}. Lithium abundance is sensitive to non-local thermodynamic equilibrium (NLTE). We computed the 3D NLTE corrections (as described in Paper II) if only 1D local thermodynamic equilibrium values were available in the original papers.\\

Our analysis is completed considering observations for Be. Be surface abundances were extracted from \citet[][]{2020ApJ...888...28B} for M67 Galactic open cluster, and from samples of field stars observed by \citet[][]{2004A&A...425.1013S,2011A&A...530A..66G,2012ApJ...746...47D,2022ApJ...941...21B}. We kept only those stars with a metallicity between [Fe/H]= +0.15 and [Fe/H] = -0.15, and with log g $\le 4.50$. We adopted the meteoritic abundance A(Be) = 1.41 by \citet[][]{2003ApJ...591.1220L} as the initial abundance of Be as done by \citet[][]{2020ApJ...888...28B}. A more recent value of A(Be) = $1.30\pm0.03$ was inferred by \citep{2009ARA&A..47..481A}, we discuss potential consequences on the results in Sect.~\ref{sect:Be}.

\begin{table*}[t!]
    \centering
    \caption{Main properties of the open clusters considered in this work, with the corresponding references for the Li and Be abundances.}
    \begin{tabular}{c|c|c|c|c|c}
         Name & Age (Myrs) & Ref Age & [Fe/H] & Ref Li & Ref Be \\
         \hline \hline 
         NGC 2420 & 2500 & I & -0.05$\pm$0.02 & I & -  \\
         M 67 (NGC 2682) & 3640 & II & -0.01$\pm$0.04 & III & IV \\
         \hline
    \end{tabular}
    \tablefoot{\\ I: \citet{2020A&A...643A.164S}; II: \citet{2019A&A...623A.108B},  III: \citet{2012A&A...541A.150P}; IV: \citet[][]{2020ApJ...888...28B}\\
}
    \label{tab:OCinfos1}
\end{table*}

\subsection{Determination of mass}

We determined the mass for the sample of eight SGB stars and KIC 4448777 corresponding to our model (see Table~\ref{tab:data}). We do not aim to find the accurate mass as done in seismic analysis but a relevant value to explore the general evolution behaviour of each star and confirm the relevance of our parametrisation for angular momentum transport. We determined the masses for the sample of RGB stars \citep[][]{2018A&A...616A..24G} using the asteroseismic scaling relations, initially proposed by \citet[][]{1995A&A...293...87K}, and updated for RGB stars as proposed by \citet[][]{2013A&A...550A.126M}, and as done by \citet[][]{2018A&A...616A..24G}. We select only stars equal or less massive than $1.55 M_{\odot}$. \\ Concerning the samples for Li and Be observations, we use masses indicated by \citet{2022ApJ...941...21B,2023ApJ...944L...5M} for their sample. As we do not have access to asteroseismic parameters for all the samples, we redetermined the masses for each star thanks to a comparison with our models including both atomic diffusion and rotation (models $_{\nu}R1$, see details in Sect.~\ref{STAREVOL}), and using a grid of eight masses (0.8-1.5 $M_{\odot}$, step of 0.1 $M_{\odot}$) at solar metallicity. We note that our models are in agreement with masses by \citet{2022ApJ...941...21B,2023ApJ...944L...5M} within the given uncertainty. 

\section{The stellar evolution code}
\label{STAREVOL}

\subsection{General assumptions}
\label{subsection:Generalities}

The models presented here are build from the conclusions of Paper I on different transport processes of chemicals and angular momentum (namely: rotation, penetrative convection, extra turbulence, extra viscosity, and atomic diffusion) during the PMS and MS. \\  
We applied the prescriptions that were identified to be the most relevant for PMS/MS low-mass stars and explored the predictions at more evolved stages until the RGB bump. First, we computed models for each of the stars described in Table~\ref{tab:data}, and investigated the required angular momentum transport efficiency. Second, we computed a grid of models in a mass range between 0.8 M$_\odot$ and 1.5 M$_\odot$ ($\delta \mathrm{M} = 0.1 \mathrm{M}_\odot$) at solar metallicity. Computations are started on the Hayashi track at the beginning of the deuterium burning phase on the PMS that we consider as the time zero of the evolution. 
In addition, we defined: 1) the zero-age main sequence (ZAMS) as the point where either the central mass fraction of hydrogen has decreased by 0.02$\%$ compared to its surface value or where the central temperature has reached $3\times10^7$K; 2) the terminal-age main sequence (TAMS) and beginning of the SGB phase as the point where the central mass fraction of hydrogen is lower than $10^{-7}$; and 3) the base of the RGB and end of the SGB as the point where the mass coordinate at the top of the H-burning shell reaches a local minimum.

\subsection{Input physics}
\label{sub:inputphysics}
We used version 4.0 of the stellar evolution code STAREVOL, as described in Paper I
\citep[for general information and previous versions, see][]{2000A&A...358..593S,2006A&A...453..261P,2009A&A...495..271D,2012A&A...543A.108L,2019A&A...631A..77A} and we refer to this work for further details and references. 
The models presented in this work were computed with the same inputs physics (equation of state, opacities, nuclear reactions, model atmosphere, and mass loss) as in Paper I.
We used the same constant value for the mixing length parameter (classical model: $\alpha_{\rm MLT}=2.1100$, models including rotation: $\alpha_{\rm MLT}=2.2236$, assuming the Schwarzschild criteria for convective stability) and the initial abundances that resulted from model calibrations on the Sun \citep[reference abundances from][]{2009ARA&A..47..481A,2018ApJ...855...15Y}.

\subsection{Chemical mixing}
\label{sub:dmicdadturb}
Atomic diffusion was implemented according to \citet{1986ApJS...61..177P} and \citet{1994ApJ...421..828T}. We did not take radiative accelerations into account. Penetrative convection from convective envelope was treated as an overshoot and computed following the formalism of \citet{2019ApJ...874...83A}. It was recently validated by asteroseismic constraints by \citet{2021A&A...656A.121A}. The depth of the overshoot was calibrated in Paper I to reproduce the Li abundance in solar-type stars in very young open clusters. This process, as formalised and calibrated for PMS/MS stars in Paper I, is no longer relevant beyond the TAMS and  phases of slow rotation. In particular, the constant mixing length parameter used along evolution is not relevant and it would lead to a non-realistic penetration depth during evolved phases of evolution due to the slowdown of the surface rotation. A complete treatment of the formalism by \citet{2019ApJ...874...83A} in the code ought to include a rotation dependence on the mixing length parameter, as well as taking into account the structural evolution along the giant phases. This approach will be undertaken in a future work. Penetrative convection is consequently stopped when reaching the TAMS. We further discuss the relevance of additional transport of chemicals along the giant phases in Sect.~\ref{sect:grilleli}. 

A core overshoot parameter of $d_{\rm over}=0.05\,H_p$, with $H_p$ as the pressure scale height, is taken into account for stars presenting a convective core along evolution (masses equal or above 1.2$M_{\odot}$ at solar metallicity).\\

Parametric turbulence is defined according to \citet{2000ApJ...529..338R} and \citet{2005ApJ...619..538R} based on the following prescription for the diffusion coefficient:
\begin{equation}
    D_{\rm{T_0}} = 400 D_{\rm{He}}(T_0)\left[\frac{\rho(T_0)}{\rho}\right]^3,
\label{eq:dturb1}
\end{equation}
where $T_0$ is a free parameter that sets the depth of the maximum efficiency of the mixing depending on the value of the atomic diffusion coefficient He ($D_{\rm{He}}$). We calibrated the efficiency of turbulence in order to reproduce the surface Li depletion in MS solar twins. We adopted $\log T_0 = 6.42$ as in Papers I (solar twins calibration). Beyond TAMS, we stopped the turbulence. 

\subsection{Angular momentum evolution and rotation-induced mixing}
\label{sub:stellarrotation}
Stellar rotation was implemented in STAREVOL as described by \citet{2016A&A...587A.105A,2019A&A...631A..77A} and Paper I. We followed the shellular rotation hypothesis developed by \citet{1992A&A...265..115Z}, \citet{1998A&A...334.1000M}, and \citet{2004A&A...425..229M} to describe the transport of angular momentum and chemicals by meridional circulation (as an advective process for angular momentum) and turbulent shear (vertical and horizontal). \\
The transport of angular momentum obeys the advection-diffusion equation:
\begin{equation}
    \rho \frac{d}{dt}(r^2 \Omega) = \frac{1}{5 r^2} \frac{\partial}{\partial r} (\rho r^4 \Omega U_2) + \frac{1}{r^2} \frac{\partial}{\partial r}\left((\nu_v+\nu_{\rm{add}}) r^4 \frac{\partial \Omega}{\partial r}\right),
    \label{eq:rot}
\end{equation}
where $\rho$, $r$, $\Omega$, $U_2$, and $\nu_v$ are the density, the radius, the angular velocity, the meridional circulation velocity, and the vertical shellular component of the turbulent viscosity, respectively. Also, $\nu_{\rm add}$ is an additional parametric viscosity as introduced by \citet{2012A&A...544L...4E,2019A&A...621A..66E} and \citet{2016A&A...589A..23S} to simulate the efficiency of a transport of angular momentum along evolution. We assumed $\nu_{\rm add}$  to be constant in time or dependent on the radial rotational shear, defined by \citet{2016A&A...589A..23S}\footnote{This scaling has been shown to be consistent with the hypothesis of a transport by the azimuthal magneto-rotational instability \citep[AMRI,][]{2018PhR...741....1R}.} as:
\begin{equation}
    \nu_{\rm{add}}(t) = \nu_{0} \times \left(\frac{\overline{\Omega}_{rad}}{\overline{\Omega}_{conv}}\right)^{\alpha}
    \label{eq:nuaddt}
,\end{equation}
where $\nu_0$ and $\alpha$ are free parameters; then, $\overline{\Omega}_{conv}$ is the mean angular velocity in the convective envelope. The envelope is assumed to rotate as a solid body and we have $\overline{\Omega}_{conv}(t) \equiv \Omega_{\rm surf} (t)$. $\overline{\Omega}_{rad}$ is the mean angular velocity in the radiative interior, given by:
\begin{equation}
    \overline{\Omega_{\rm rad}} = \frac{\int_{M_{\rm TCC}}^{M_{\rm BCE}} r^2 \Omega \, dm}{\int_{M_{\rm TCC}}^{M_{\rm BCE}} r^2 dm},
\end{equation}
with $M_{\rm TCC}$ as the mass coordinate of the top of the convective core and $M_{\rm BCE}$ as the mass coordinate at the base of the convective envelope. As noted in the reference papers, we do not expect that the missing process for angular momentum transport would be constant in time and/or in space, but this approach allows us to simulate the required transport efficiency along evolution and model the correct rotation structure.\\

We explored models that include prescriptions for turbulence shear, referred to as R1 in Paper I; R1 includes prescriptions from \citet{2018A&A...620A..22M} and \citet{1992A&A...265..115Z} for the horizontal diffusivity, $D_h$, and the vertical diffusivity, $D_v$, respectively. The detailed expressions of the two turbulent diffusion coefficients can be found in Appendix~B of Paper I. We used the same parameters as in Paper I for the extraction of angular momentum at the stellar surface due to magnetised winds (m=0.22, p=2.1, $\chi=14$, $K=7.5\times10^{30}$ erg, unless otherwise indicated) according to the \citet{Matt2015} formalism. Models were computed for a value of the initial rotation period on the PMS of 4.5 days, which are referred as the median rotating model in \citet[][]{2019A&A...631A..77A}, and Papers I and II. The disc coupling timescale was set at $\tau_{\rm disc}$ = 5 Myrs, in agreement with \citet{2015A&A...577A..98G} and \cite{2019A&A...631A..77A}.

\section{Angular momentum transport in SGB and RGB phases}
\label{sect:adjust}
In this section, we explore the observational behaviour of the core rotation rate along evolution, and we compare it to the predictions of rotation models $_{\nu}R1$ (including a constant additional viscosity) and $_{\nu(t)}R1$ (including a time-dependent additional viscosity). We investigated the required efficiency for the transport of angular momentum in view of the observational constraints from asteroseismology.  

\begin{SCfigure*}[1.0][t!]
         \includegraphics [width=110mm]{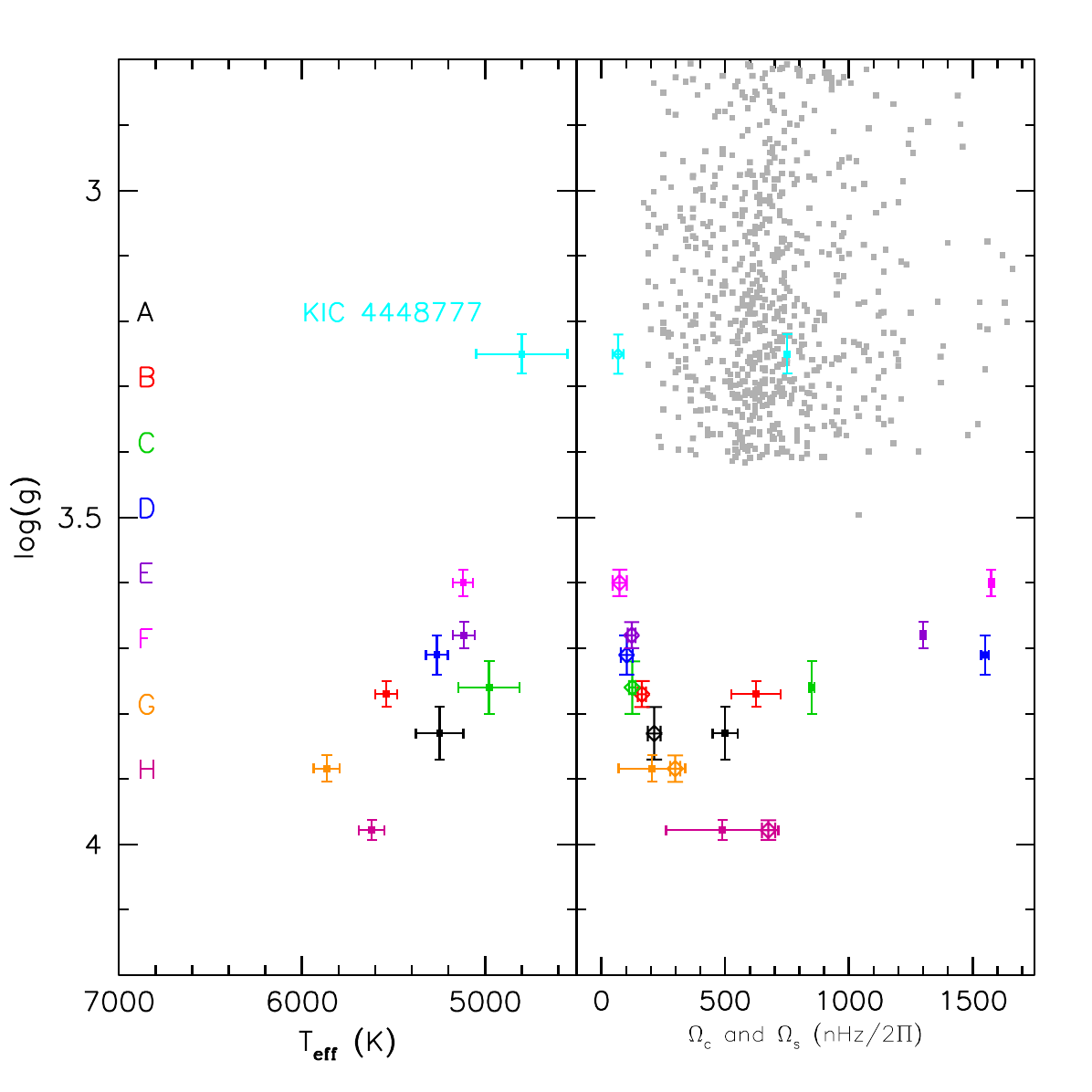}
         \caption{Observational constraints for the sample of SGB and RGB stars. Left panel: Kiel diagram. Individual squares represent the nine stars of Tab.~\ref{tab:data} (colour-coded). Right panel: Core rotation rates (filled squared) and surface rotation rates (open diamonds) of the sample of evolved stars Tab.~\ref{tab:data}. 
         Parameters for the eight subgiants come from \citet{2014A&A...564A..27D,2020A&A...641A.117D}. Parameters for KIC 4448777 come from \citet{2016ApJ...817...65D}.}
         \label{fig:obs}
 \end{SCfigure*}

\subsection{Internal and surface rotation in evolved stars}
\label{sect:observations}
The evolution of the core rotation rate in evolved stars is observed thanks to the asteroseismic constraints as described in Sect.~\ref{sec:obs}. Figure~\ref{fig:obs} shows the Kiel diagram on the left panel, where the nine stars of our sample are represented (colour-coded), and the core (filled squares) and surface (open diamonds) rotation rates as a function of surface gravity on the right panel. In addition, the sample of RGB stars from \citet[][]{2018A&A...616A..24G} is represented in grey squares for the core rotation rate only. 

As the stars evolve on the SGB and RGB, the surface rotation rate is progressively slowing down, while the convective envelope is expanding, eventually resulting in slow surface rotations for each star. In parallel, the contraction of the core drives an acceleration of the core rotation rate, but the observed behaviour is more complex and call for the action of a transport of angular momentum.\\
There are three observables to consider to understand the transport efficiency from these observations: a) the mass of the star; b) the metallicity of the star; and c) the evolutionary state of the star. \\ As observed in previous studies, the transport efficiency has been shown to increase with the mass and decrease with the metallicity during post-MS phases -- or, in other words, to increase with the temperature \citep{2014A&A...564A..27D,2020A&A...641A.117D,2017A&A...599A..18E,2019A&A...621A..66E,2022A&A...663A.180M,2023arXiv230207811M}. As for evolution, the transport of angular momentum is not trivial. We can distinguish three main steps in the angular momentum transport evolution observing the present sample of nine stars:\\
    First, efficient transport is involved during the MS and beginning of SGB. It leads to a coupling between the core and surface rotation as illustrated by the Sun and by the young SGB stars G and H that exhibit an almost uniform rotational profile. This suggests a continuity with respect to the same process from the MS to the beginning of the SGB phase.\\
    Second, along the SGB phase, the transport efficiency is decreasing. The process driving the transport of angular momentum may become inefficient. The evolution along the SGB phase leads to structural changes and a natural increase in the core rotation rate. We observe an increase in the differential rotation that arises for stars A to F, and especially for old SGB stars D, E, and F.\\
    Third, when reaching the RGB phase, the differential rotation tends to stay constant at an intermediate level compared to SGB stars, as observed from the sample by \citet[][]{2018A&A...616A..24G}. A stronger transport efficiency is then required to keep an almost constant core rotation rate, and consequently an increased efficiency compared to the SGB phase. This last step suggests the onset of a new transport process taking place along the RGB phase.\\  
As described in Sect.~\ref{section:introduction}, the identification of the involved transport processes is not trivial. Internal gravity waves (hereafter, IGW) could be a promising candidate in that context \citep[][]{2008A&A...482..597T,2014MNRAS.444..102K,2017A&A...605A..31P} as we would expect them to efficiently transport angular momentum along the MS phase before becoming inefficient along the SGB phase (as observed in our second point). However, IGW have been seen to damp before reaching the internal layers along the RGB phase and cannot drive the required transport. Magnetic fields and instabilities \citep[e.g.][]{2019MNRAS.485.3661F} or mixed-modes themselves \citep[][]{2015A&A...579A..31B} have been considered, as well as other promising candidates, to efficiently transport angular momentum along evolution. In particular, \citet[][]{2022NatAs.tmp..119E} showed the relevance of the Tayler-Spruit instability in reproducing the Sun. However, \citet[][]{2019A&A...631L...6E,2020A&A...641A.117D} showed that it is challenging to reproduce both SGB and RGB phases (and MS) with a single process. A combination of several processes taking place at different temperatures and evolutionary steps is consequently supported, especially in the context of the sharp transition observed in the SGB phase. 

\subsection{Model predictions for internal rotation: constant extra viscosity}
\begin{figure}[!ht]
         \center
         \includegraphics [width=90mm]{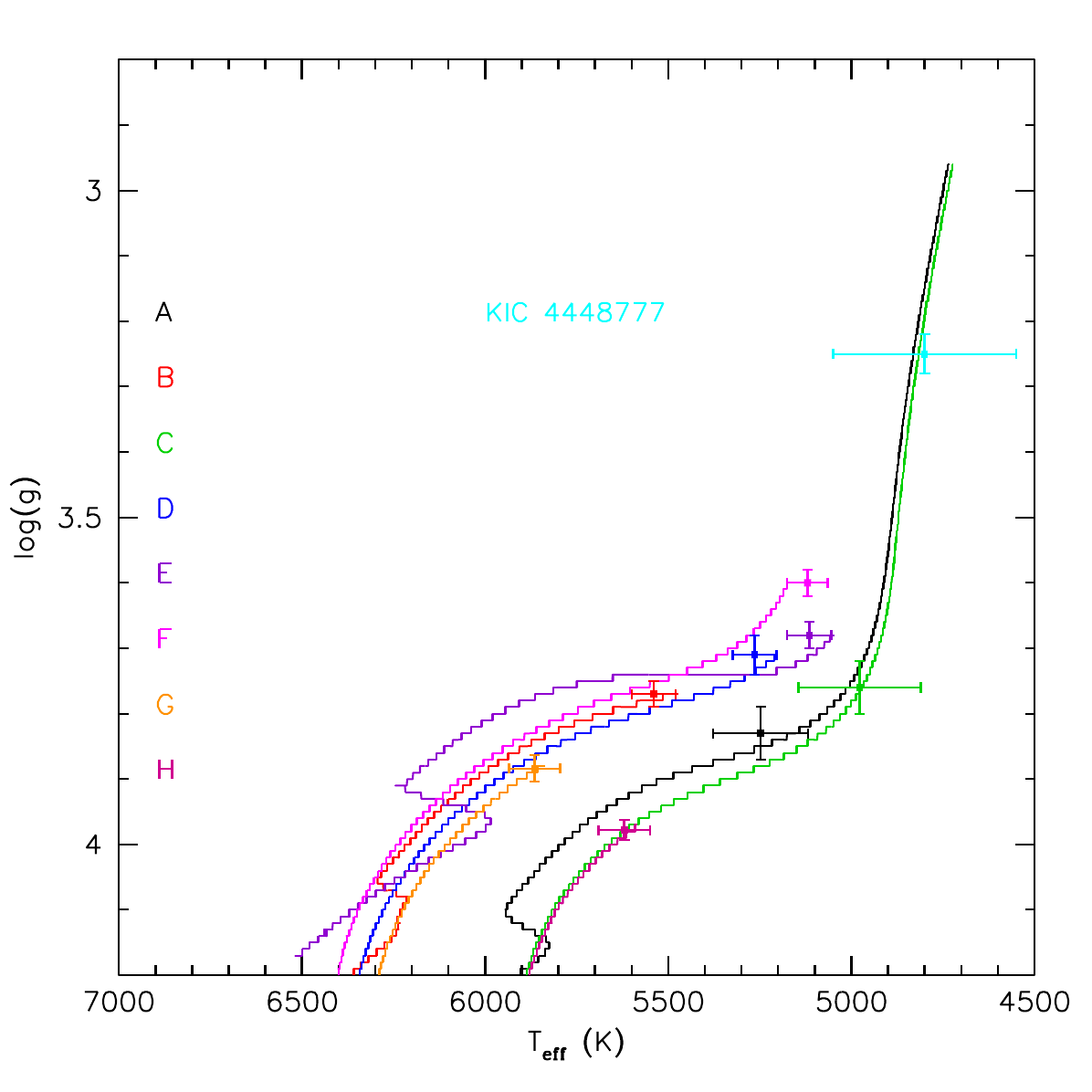}
         \caption{Same as left panel of Fig.~\ref{fig:obs} where the evolutionary tracks of the adjusted stellar models, $_{\nu}R1,$ for each SGB stars are shown (see Table~\ref{tab:data}).}
         \label{fig:hrdadjust}
 \end{figure}

We now model the nine stars from Table~\ref{tab:data}, adjusting the mass for each star in order to reproduce the effective temperature, $T_{\rm eff}$, and surface gravity log(g) within the errors, as we do not aim to reproduce exactly these stars -- but, rather, their overall evolutionary behaviour. The evolutionary tracks for each star are shown in the Kiel diagram of Fig.~\ref{fig:hrdadjust} \footnote{Little steps appear on the tracks of Fig.~\ref{fig:hrdadjust} due to the surface gravity model outputs accuracy. They are too small to expect any impact on the result of the paper.}. We modelled the star KIC 4448777 using the same mass and metallicity as star C, since it shares similar values for the metallicity and seismic mass. The track for star A at the same metallicity and a slightly higher mass is also extended in order to explore the predictions from SGB to RGB phases.    
 
\begin{SCfigure*}[1.0][t]
         \includegraphics [width=110mm]{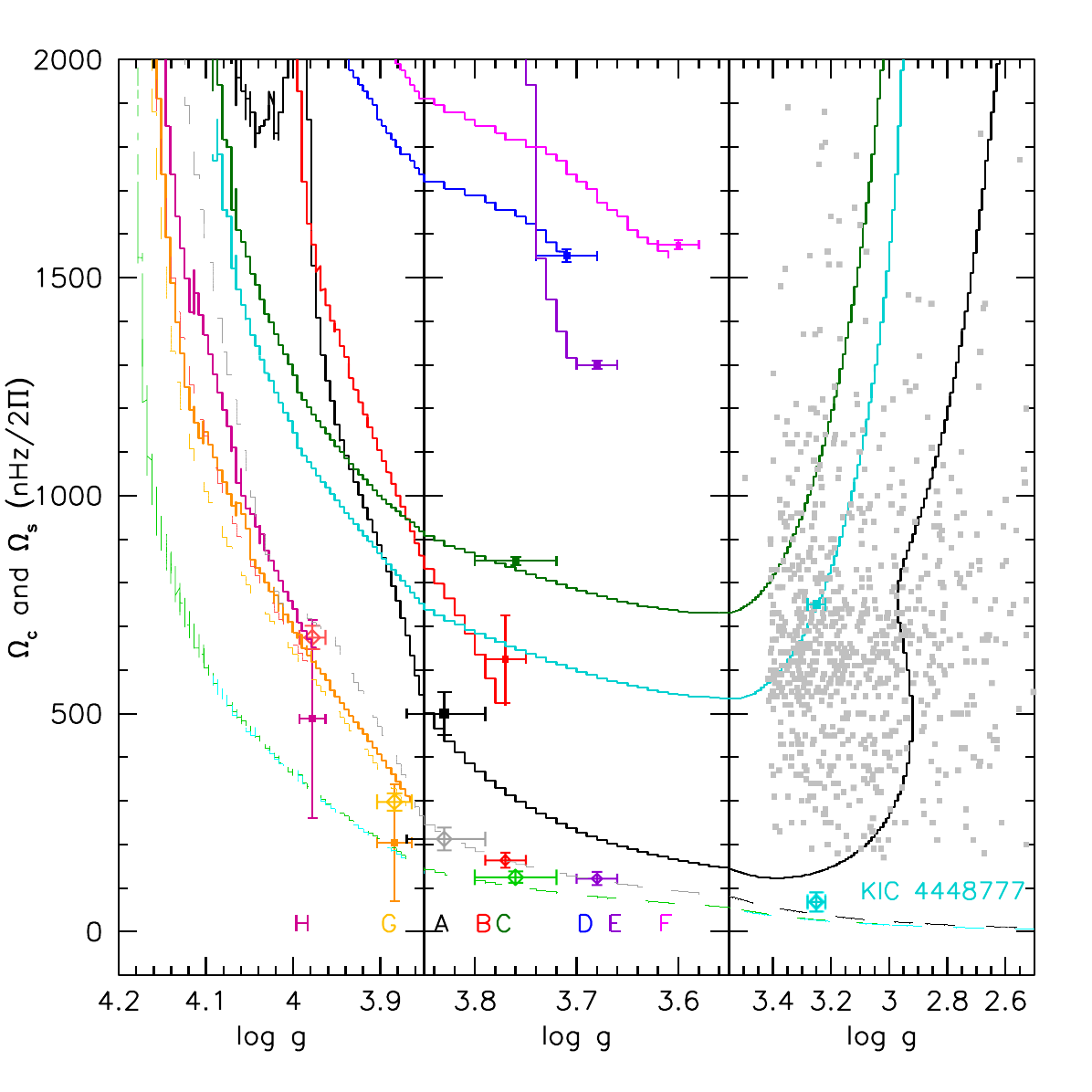}
         \caption{Core (full line) and surface (dashed-line) rotation rates as a function of surface gravity. Comparison to $_{\nu}R1$ model predictions with values of extra viscosity $\nu_{\rm add}$ adjusted for each star to reproduce the core rotation rate (see Table~\ref{tab:datanu}). The three panels discriminate the three evolutionary steps described in Sect.~\ref{sect:observations}. (Surface rotation rates are shown only for stars A, C, G, H, and KIC 4448777 in lighter colour shade for sake of clarity).}
         \label{fig:omc_sgb}
 \end{SCfigure*}

In this section, we estimate the required efficiencies for the additional angular momentum transport adjusting a constant value of the extra viscosity, $\nu_{\rm add}$, for each star to reproduce the internal rotation rate inferred by asteroseismology. Also, we adjusted the parameter K for the magnetic braking as defined by \citet[][]{Matt2015} to adjust the surface rotation rate (if needed). We note that here we consider an initial rotation velocity of 4.5 days. Figure~\ref{fig:omc_sgb} shows the predicted core rotation rates obtained for the nine stars of Table~\ref{tab:data} and the surface rotation rate for stars A, C, G, H, and KIC 4448777. We divided the plot in three boxes corresponding to the three steps observed and described in Sect.~\ref{sect:observations}.
 
 Table~\ref{tab:datanu} gives the adjusted values of the additional viscosity for each star compared to the ones found in previous studies from \citet{2019A&A...621A..66E} and \citet[][]{2020A&A...641A.117D}. 
 We also indicate in Table~\ref{tab:datanu} the values used for parameter K. In particular, we adjusted parameter K for A, B, and E massive stars of our sample, as well as the young SGB star H, which exhibits an almost solid-body rotation with a relatively high surface rotation rate. 
 \begin{table*}[t]
    \centering
    \caption{Additional viscosities adjusted for each stars of Table~\ref{tab:data} and values from reference papers. Stars are grouped in three different groups corresponding to the three evolutionary steps described in Sect.~\ref{sect:observations}.}
    \begin{tabular}{c|c|c|c|c}
         Star & Magnetic braking & $\nu_{\rm add}$ & Magnetic braking & $\nu_{\rm add}$ \\
         & K ($10^{30}$ erg) - This work & This work & K ($10^{30}$ erg) - Literature & Literature \\
         \hline \hline
         G & 7.5 & $>5.0\times10^5$ & None / 6.3 & $>1.5\times10^5$ / $>2.5\times10^5$ \\
         H & 0.75 & $>1.0\times10^5$ & None / 6.3 & $>5.0\times10^4$ / $>8.0\times10^4$ \\
         \hline 
         B & 10.5 & $5.0\times10^4$ & None & $(1.7 \pm 0.6)\times10^4$ \\
         E & 30 & $4.35\times10^4$ & None & $(1.0 \pm 0.2)\times10^4$ \\
         A & 1.5 & $3.0\times10^4$ & None & $(1.5 \pm 0.5)\times10^4$ \\
         D & 7.5 & $8.75\times10^3$ & None & $(6.0 \pm 2.0)\times10^3$  \\
         F & 7.5 & $8.75\times10^3$ & None & $(4.0 \pm 2.0)\times10^3$  \\
         C & 7.5 & $7.25\times10^3$ & None & $(6.0 \pm 1.5)\times10^3$ \\
         \hline
         KIC 4448777 & 7.5 & $8.5\times10^3$ & - & - \\
         \hline
    \end{tabular}
    \label{tab:datanu}
    \tablefoot{References. Stars A to F: \citet[][]{2019A&A...621A..66E}, Stars G and H: \citet[][]{2020A&A...641A.117D}}
\end{table*}

In the present study, the strongest viscosities are obtained for the young SGB G and H stars ($> 5.0\times10^5$ and $> 1.0\times10^5$ $\rm cm.s^{-1}$, respectively). They correspond to nearly solid rotation models and they confirm that a strong coupling is required along the early part of SGB phase, independently of the mass, the metallicity, and the magnetic braking, in agreement with the hypothesis of a sharp transition during the SGB phase before a decreasing efficiency of the transport. Concerning the older SGB stars, the weakest viscosity is obtained for the 1.15$M_{\odot}$ SGB metal-rich C star ($7.25\times10^3$ $c\rm m.s^{-1}$), and the strongest viscosity is obtained for the metal-poor 1.2$M_{\odot}$ B star ($5.0\times10^4$ $c\rm m.s^{-1}$). Considering stars A and C that share a close metallicity, the more massive and hotter A star requires a higher viscosity than C star ($3.0\times10^4$ $\rm cm.s^{-1}$ and $7.25\times10^3$ $\rm cm.s^{-1}$, respectively). Considering stars A and B, which have the same mass, the metal-poor B star requires a higher transport efficiency than  star A  ($5.0\times10^4$ $\rm cm.s^{-1}$ and $3.0\times10^4$ $\rm cm.s^{-1}$, respectively). In other words, the required viscosity increases with an increasing temperature. \\

We note that the available sample is composed of stars with a relatively small range of metallicities and masses. A larger sample would help to make a definitive conclusion on the trend. However, the range is already significant for low-mass stars and our results are in good agreement with the trends and results from previous studies by \citet[][]{2014A&A...564A..27D,2020A&A...641A.117D,2017A&A...599A..18E,2019A&A...621A..66E,2022A&A...663A.180M}. It supports the robustness of this observation, especially considering differences on the codes, shear turbulence prescriptions, initial rotation and magnetic braking efficiency. However, we note  that the initial rotation velocity should not significantly affect  the results as observed by \citet[][particularly visible in their Fig. 9]{2019A&A...621A..66E} for late SGB stars and confirmed for RGB stars by \citet[][]{2022A&A...663A.180M}, who computed models with periods between 2 and 50 days and did not observe a different behaviour in the core rotation rate evolution. However, the case of the young SGB stars (G and H) is not clear, the rotational history may be important at this phase close to the TAMS and may lead to some degeneracy. In all our models, we considered a magnetic braking contrary to \citet[][]{2019A&A...621A..66E} who also considered a slow rotation velocity. In the work by \citet[][]{2020A&A...641A.117D} on stars G and H, it is shown that the values obtained for the viscosity are smaller without magnetic braking (see Table~\ref{tab:datanu}). This trend is confirmed by the comparison of our results with \citet[][]{2019A&A...621A..66E}, where we obtained the same order of magnitude despite slightly higher values for each stars.

Interestingly, the efficiency of the extra viscosity we adjusted for each SGB stars (except young SGB stars G and H) is smaller or close to the value of $3.5\times10^4$ $\rm cm.s^{-1}$ we obtained for the 1.0$M_{\odot}$ and solar metallicity MS solar-type evolution in Paper I, with the same input physics. It is only the more massive B and E stars that require stronger efficiencies. We also computed the required efficiency to reproduce the KIC 4448777 RGB star and estimated a value of $8.5\times10^3$ $\rm cm.s^{-1}$. It is on the same order as the low-mass SGB stars C, D, and F, and supports the notion that the efficiency of the transport should not decrease further after reaching the RGB phase. In addition, we observe a strong increase in the core rotation rate predicted by the models along RGB phase, which is in contradiction to the almost constant evolution observed in the RGB phase as discussed by \citet[][]{2018A&A...616A..24G}, and independently of mass (see for instance their Figs. 11 and 12). A stronger efficiency in the transport of angular momentum is then required when reaching the RGB phase.

These results confirm that a constant efficiency for the extra viscosity is not relevant throughout evolution. It is also evident that the transport of angular momentum transport is required to vary over time, potentially involving different transport processes from the MS/beginning of the SGB to the end of the SGB/RGB phases. In order to simulate the effect of one or several transport processes involved along evolution, we now explore a rotation-dependent additional viscosity as described by \citet[][]{2016A&A...589A..23S} and related to AMRI, recently tested and validated by \citet{2023arXiv230207811M}, and where we consider a mass dependency as well .

\subsection{Model predictions for internal rotation: rotation-dependent extra viscosity}
\label{sect:nuaddt}
We go on to explore the model predictions considering an additional viscosity dependent on time, following Eq.~\ref{eq:nuaddt}. In the previous section, we focus on a sample of SGB stars in order to confirm the behaviour of the core rotation rate evolution with our models. We now extend our analysis to a grid with a mass range from 0.8$M_{\odot}$ to 1.5$M_{\odot}$ (0.1$M_{\odot}$ step) at solar metallicity, as defined previously (see Sect.~\ref{subsection:Generalities}). 

\begin{figure*}[t!]
         \center
         \includegraphics [width=150mm]{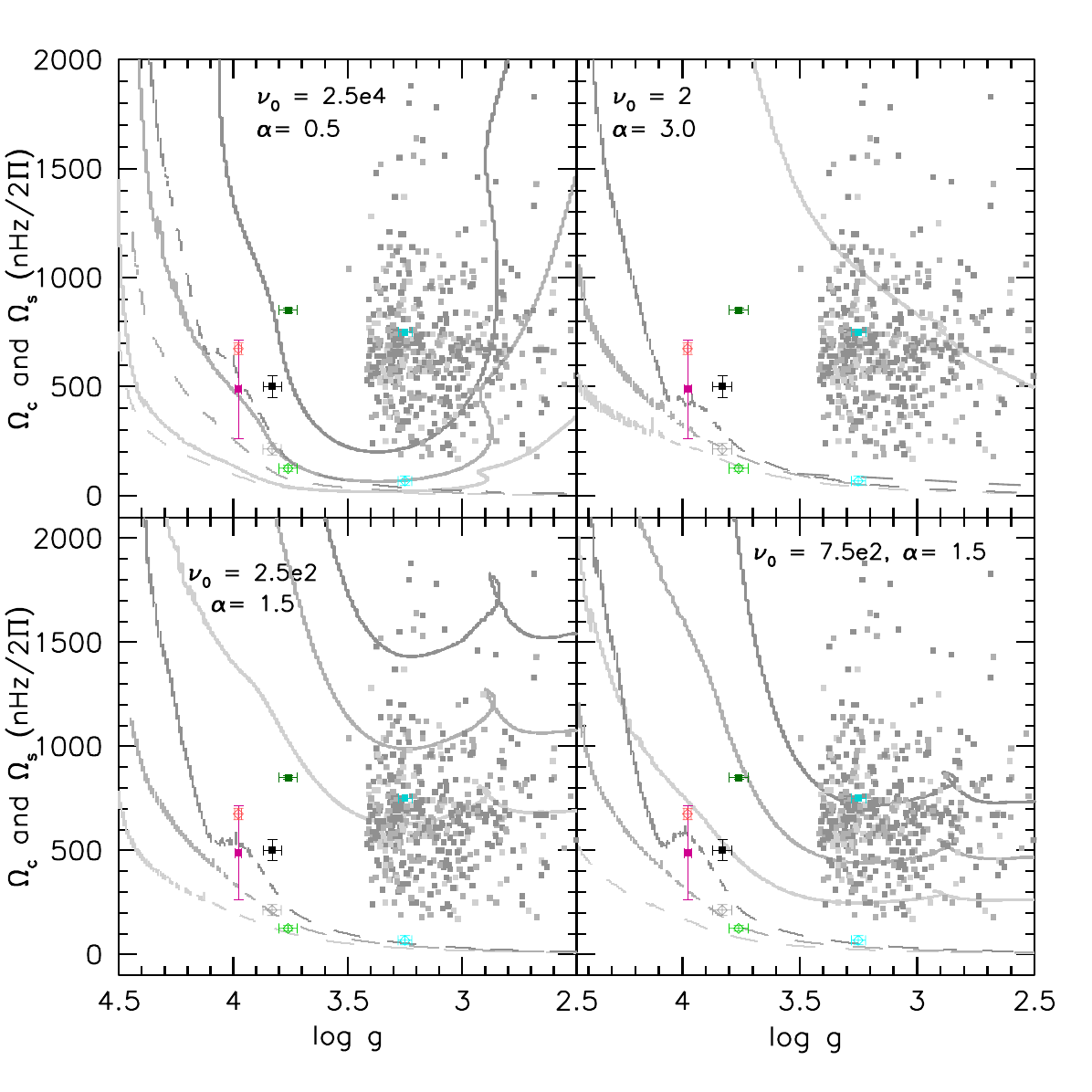}
         \caption{Core (full line) and surface (dashed-line) rotation rates for 1.0 (light grey), 1.1 (medium grey), and 1.2 (dark grey) $M_{\odot}$ models. Different sets of coefficients for the extra-viscosity $\nu_{\rm add}(t)$ are explored in each panel: top-left panel: $\nu_0=2.5\times10^4$/$\alpha=0.5$, top-right panel: $\nu_0=2$/$\alpha=3.0$, bottom-left panel: $\nu_0=2.5\times10^2$/$\alpha=1.5$, and bottom-right panel: $\nu_0=7.5\times10^2$/$\alpha=1.5$. A, C, and H SGB stars, and KIC 4448777 are indicated with the same colour code as in previous figures. RGB stars from \citet{2018A&A...616A..24G} are colour-coded as the models predictions according to their determined seismic mass.
         }
         \label{fig:omc_grille}
 \end{figure*}

We explored the relevance of the time-dependent viscosity and its mass dependence at solar metallicity. Figure~\ref{fig:omc_grille} shows the core rotation rate in light grey, medium grey, and dark grey, for models of 1.0, 1.1, and 1.2 $M_{\odot}$, respectively. The models were computed for four different sets of coefficients $\nu_0$ and $\alpha$ with the same magnetic braking K parameter (adjusted for the Sun). 

The top-left panel shows the result with the set adjusted in Paper I for MS solar-type stars ($\nu_0 = 2.5\times10^4$ and $\alpha=0.5$) and similar to the results obtained with a constant $\nu_{\rm add} = 3.5\times10^4$ $\rm cm^2.s^{-1}$ (not shown). As expected from the smaller values obtained in Table~\ref{tab:datanu}, the transport is overly efficient, leading to slow core rotation during the SGB and RGB phases. The general behaviour is the same as that seen with a constant viscosity. On the other hand, in top right-panel of Fig.~\ref{fig:omc_grille}, we show the set ($\nu_0 = 2$ and $\alpha = 3$) adjusted by \citet[][]{2016A&A...589A..23S} for a 1.25$M_{\odot}$ for the SGB and RGB phases with solid rotation imposed either until TAMS or until 1 Gyr after the TAMS. We note that in our case, we considered the extra viscosity all along evolution as well as a magnetic braking adjusted on the Sun. As shown by \citet[][]{2016A&A...589A..23S}, it is possible to reproduce the trend of a decreasing core rotation rate fixing a strong dependence on the differential rotation with $\alpha=3$. In this case, the high dependency on the differential rotation leads to a decreasing core rotation rate in both the SGB and RGB phases. The transport efficiency is weaker than the first case, and because we did not impose solid rotation during the MS, the core rotation rate remained too swift, compared to the observations. In order to adjust the transport along all the whole evolution and to reproduce the almost flat core rotation rate along RGB phase, intermediate values of coefficients are required. In the bottom panels of Fig.~\ref{fig:omc_grille}, we show predictions obtained from models with two different values of $\nu_0$ and with $\alpha = 1.5$, following the results of \citet[][]{2016A&A...589A..23S}. As the star evolves from the SGB phase to the RGB phase, the expansion of the star in parallel to the core contraction drives an increasing efficiency of the additional viscosity by the way of the ratio $\frac{\overline{\Omega_{rad}}}{\overline{\Omega_{conv}}}^{\alpha}$. An almost uniform evolution of the core rotation rate is corresponding to a value of $\alpha \approx 1.5$. This value is in good agreement with the value of $\alpha=2$ obtained by \citet[][]{2016A&A...589A..23S} and \citet{2023arXiv230207811M} for a uniform core rotation rate in the RGB phase and is retained in the following models. \\
The value of $\nu_0$ that determines the order of efficiency of the transport is not trivial and ought to depend on the temperature as seen previously. The bottom panels of Fig.~\ref{fig:omc_grille} show model predictions for values of $\nu_0 = 2.5\times10^2$ (left) and $\nu_0 = 7.5\times10^2$ (right). In both cases, the decreasing core rotation rate along SGB phase and uniform evolution along RGB phase are predicted. In addition, a mass dependence is shown. As in \citet{2022A&A...663A.180M}, we determined an average value of $\nu_0$ for each mass of our grid, where the models predict a core rotation rate ($\Omega_c/2\Pi$) between 500 and 800 nHz in order to reproduce the bulk of RGB stars from \citep[][]{2018A&A...616A..24G}. Adjusted values for each mass are reported in Table~\ref{tab:datanu0} (we note that we did not consider solid rotation in our models and that we include magnetic braking). Value of $\nu_0$ should increase with increasing mass to reproduce the bulk of RGB stars. In the same figure (Fig.~\ref{fig:omc_grille}), we indicated SGB stars A, C, H, and KIC 4448777 as references.

\begin{figure}[t]
         \center
         \includegraphics [width=90mm]{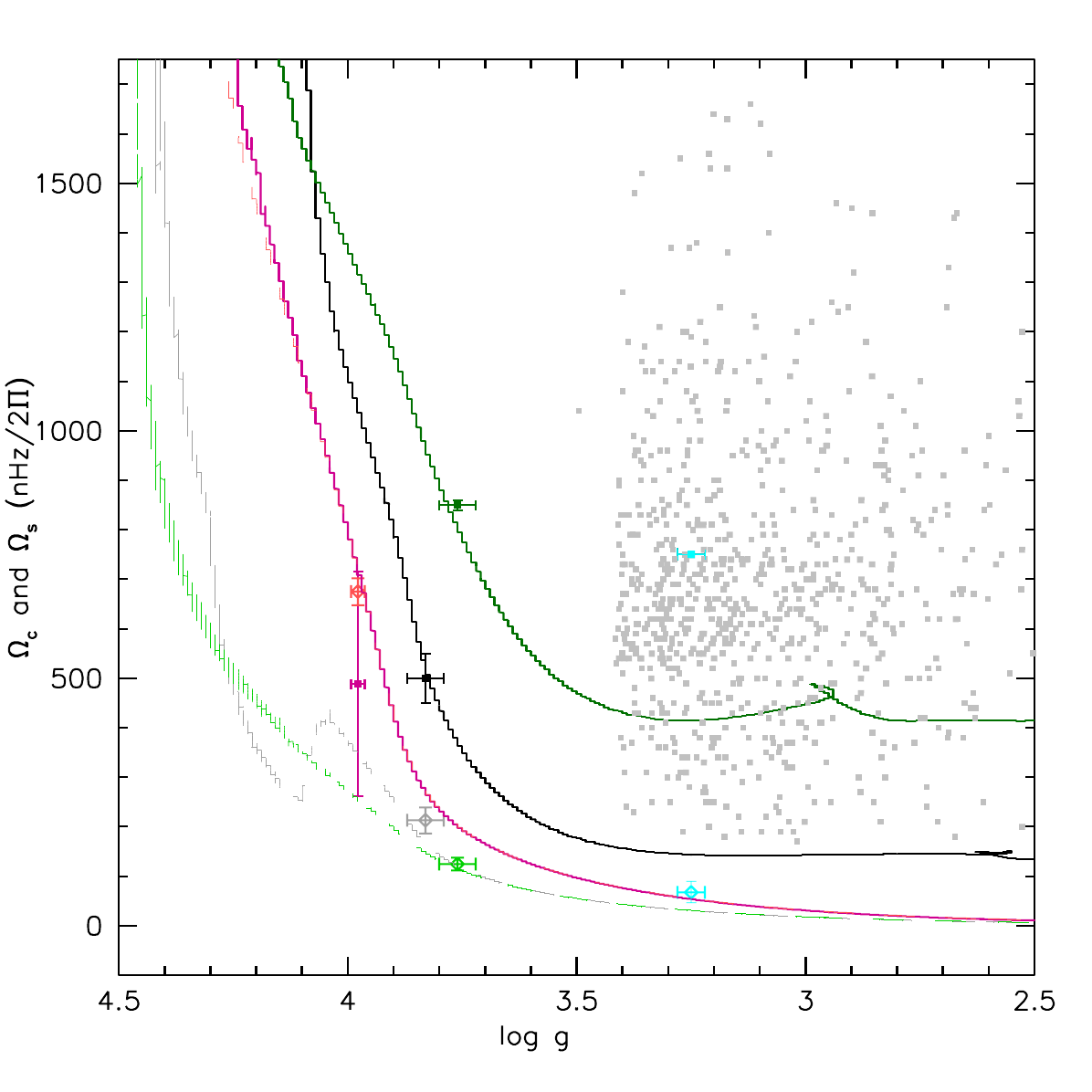}
         \caption{Same as Fig.~\ref{fig:omc_sgb} but for models $_{\nu(t)}R1$ of stars A (black), C (green), and H (purple), including a time-dependent extra viscosity with $\nu_{0.A}=5\times10^3$, $\nu_{0.C}=7.5\times10^2$, $\nu_{0.H}=10^5$, and $\alpha = 1.5$ (see Eq.~\ref{eq:nuaddt}).}
         \label{fig:omc_sgb_rgb}
 \end{figure}

We carried out the same exercise with dedicated models for the sample of eight SGB stars and KIC 4448777 at different masses and metallicities, and we report corresponding values of $\nu_0$ in Table~\ref{tab:datanu0}. We ordered the stars first by function of their mass and then by function of their metallicity [Fe/H]. Figure~\ref{fig:omc_sgb_rgb} shows the core rotation rate for models of stars A, C, and H, including the time-dependent extra viscosity. We obtained the same trend as observed when including a constant value $\nu_{\rm add}$. In other words, for a higher temperature (higher mass and/or a smaller metallicity), stars are more compact and a higher value of $\nu_0$ is required to achieve the same coupling. \\ 
 
The modified model predicts a better agreement with the core rotation rate behaviour from the SGB to the RGB phase. For each star, our models predict a relevant behaviour from the SGB to the RGB phase, both for the surface and the core rotation rates. However, in order to reproduce the almost-solid body rotation of star G and H with the same formalism, we need to impose a strong transport that results in a low core rotation velocity along the RGB that is not observed. Our formalism is consequently not adapted to these transition as seen before. It implies a sharp change of transport efficiency that cannot be reproduced by a continuous transport as done in our models, or only if imposing almost solid body rotation before to release it at a point of the SGB phase. This was done, for instance, by \citet[][]{2016A&A...589A..23S}. 

\begin{table}[t]
    \centering
    \caption{Adjusted values of $\nu_0$ for our grid and for each stars of Table~\ref{tab:data}.}
    \begin{tabular}{c|c|c|c}
         Model-Star & Mass, $M_{\odot}$ & [Fe/H] & $\nu_0$ \\
         \hline \hline
         $_{\nu(t)}R1$ & 0.80 & 0.00 & $5.0\times10^1$ \\
         $_{\nu(t)}R1$ & 0.90 & 0.00 & $1.0\times10^2$ \\
         $_{\nu(t)}R1$ & 1.00 & 0.00 & $2.5\times10^2$ \\
         $_{\nu(t)}R1$ & 1.10 & 0.00 & $5.0\times10^2$ \\
         $_{\nu(t)}R1$ & 1.20 & 0.00 & $7.5\times10^2$ \\
         $_{\nu(t)}R1$ & 1.30 & 0.00 & $1.2\times10^3$\\
         $_{\nu(t)}R1$ & 1.40 & 0.00 & $5.0\times10^3$ \\
         $_{\nu(t)}R1$ & 1.50 & 0.00 & $1.0\times10^4$\\
         \hline \hline
         H & 1.10 & +0.14 & $1.0\times10^5$ $^*$ \\
         G & 1.12 & -0.17 & $5.0\times10^5$ \\
         \hline
         F & 1.10 & -0.40 & $3.0\times10^2$ $^*$ \\
         C & 1.15 & +0.25 & $7.5\times10^2$ \\
         D & 1.15 & -0.15 & $2.0\times10^3$ \\
         A & 1.20 & +0.25 & $5.0\times10^3$ \\
         B & 1.20 & -0.09 & $1.0\times10^4$ $^*$ \\
         E & 1.50 & +0.41 & $1.0\times10^4$ $^*$ \\
         \hline
         KIC 4448777 & 1.15 & +0.25 & $3.5\times10^2$ \\
    \end{tabular}
    \label{tab:datanu0}
    \tablefoot{Stars are ordered by evolutionary state: Early SGB stars G and H, SGB stars A to F and RGB star KIC 4448777, then by increasing mass, and, finally, by decreasing metallicity. \\
    * $K \ne K_{\odot}$}
\end{table}

To summarise, the approach we used allows to simply simulate a varying transport of angular momentum along evolution reproducing well the evolution behaviour of stars along SGB and RGB phases as well as the mass trend. However, the same formalism cannot reproduce the sharp transition during the SGB phase. This suggests that two different processes are involved before and beyond this point or the need for the activation or deactivation of a single process along the evolution. In the first case, IGW could play a role but not in the second case as IGW are inefficient to transport angular momentum beyond MS/young SGB phases. On the other hand, magnetic instabilities are a promising explanation in both cases, as already discussed by \citet[][]{2020A&A...641A.117D}, and shown by \citet{2016A&A...589A..23S,2022NatAs.tmp..119E} for instance. Nevertheless, the present formalism allows us to reproduce the general rotational evolution profile from MS to SGB and RGB stars, as already done by \citet{2022A&A...663A.180M,2023arXiv230207811M}, but based on a different stellar evolution code with different initial parameters.

\section{Model predictions for chemical evolution: Lithium and beryllium}
\label{sect:grilleli}
In the previous section, we describe the construction of models that reproduce the rotation evolution from the MS to the SGB and RGB phases. We go on to explore, under this structural constraints, the transport of chemicals and especially of lithium along evolution. As observed for the case of the MS phase (see for instance Paper I and Paper II), the efficiency of the angular momentum transport can lead to important variations in the case of light elements and ought to be taken into account. 

\subsection{Surface Li abundance along SGB and RGB phases}
\label{sect:Lirgb}
\begin{figure}[t]
         \center
         \includegraphics [width=90mm]{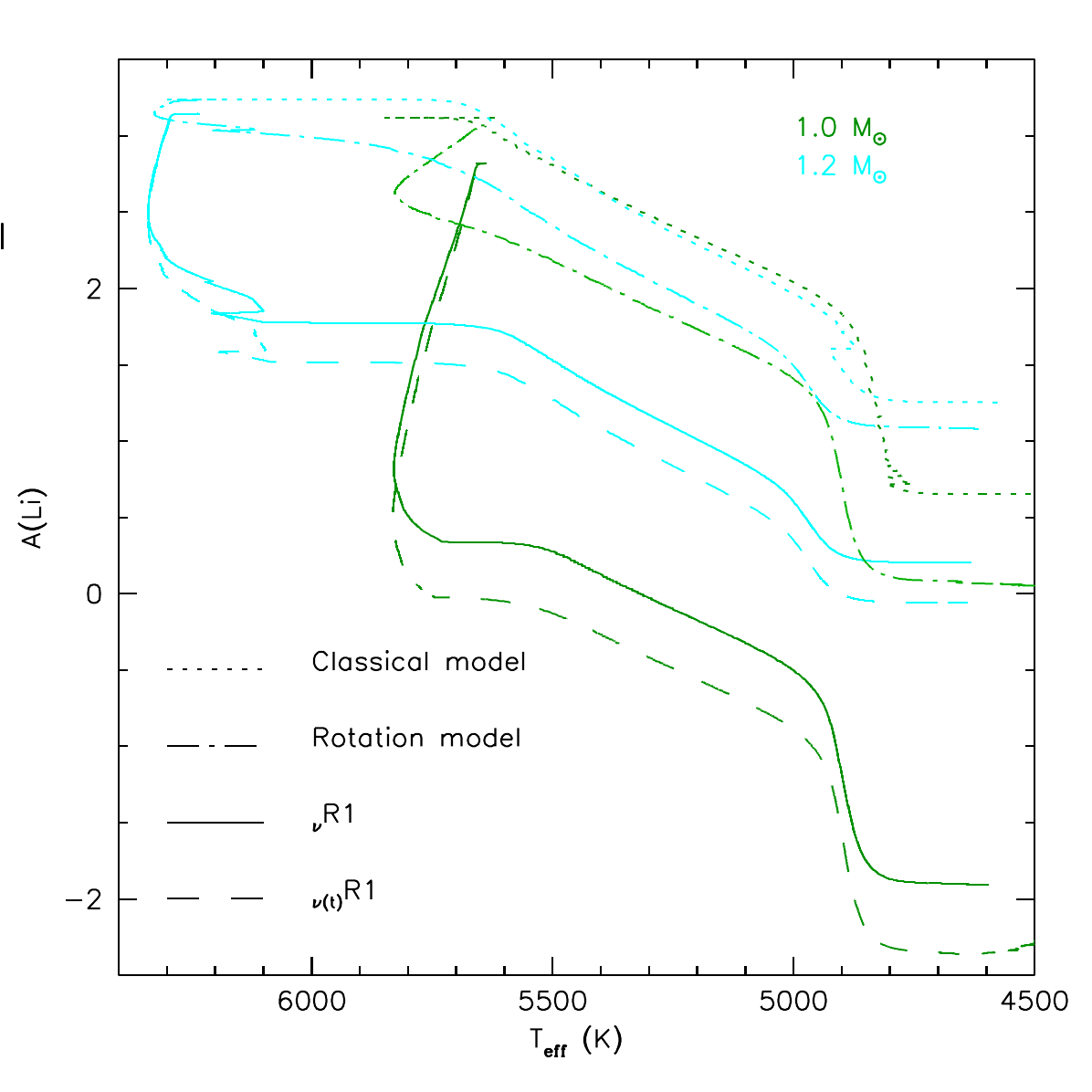}
         \caption{Evolution of Li surface abundance as a function of effective temperature for 1.0 and 1.2 $M_{\odot}$ models (colour-coded) and solar metallicity. Classical model, rotation model, $_{\nu}R1$ model, and $_{\nu(t)}R1$ model are represented in dotted, dashed-dotted, full, and dashed-line, respectively (see details in text). 
         }
         \label{fig:LiTeffproc}
 \end{figure}

We used the same grid as described in Sect.~\ref{sect:nuaddt} to explore the consequences on evolution of Li surface abundance,  focusing on the SGB and RGB phases in particular. \\ Figure \ref{fig:LiTeffproc} shows the surface Li evolution predictions as a function of effective temperature, for 1.0$M_{\odot}$ (green) and 1.2$M_{\odot}$ (cyan) stars at solar metallicity. Models predictions are given for four cases: 1) classical model: only convective transport; 2) rotation model: convection, atomic diffusion, meridional circulation, turbulence shear; 3) $_{\nu}R1$: rotation model + penetrative convection + parametric turbulence + constant extra viscosity of $\nu_{\rm add} = 3.5\times10^4$ $\rm cm^2.s^{-1}$, as defined in Paper I; and 4) $_{\nu(t)}R1$: rotation model + penetrative convection + parametric turbulence + modified time-dependent extra viscosity with $\alpha=1.5$ and the $\nu_0$ values given in Table~\ref{tab:datanu0}.

The behaviour of the different models for the Li-depletion along PMS/MS phases was largely described in Papers I and II. Mainly, our models predict a smaller amount of Li reaching the TAMS ($\approx$ 5'700 K for the $1.0 M_{\odot}$) due to the effect of the rotational mixing and also due to the effect of penetrative convection (mainly effective along PMS), parametric turbulence (mainly effective along MS), and the indirect effect of the additional viscosity. We recall that when reaching the SGB phase, we stopped these two processes (as explained in Sect.~\ref{sub:dmicdadturb}) and the predicted depletion beyond this point is due to the first dredge-up dilution and rotational mixing only. For the $1.0 M_{\odot}$, with the classical model as a reference, we predict at the TAMS a difference of about -0.5 dex for the rotational model, and -2.3 dex for the $_{\nu}R1$ model including the additional transports. The update of the time-mass dependent viscosity impacts indirectly the chemical transport and results in a stronger depletion for each model at the TAMS (at about -2.6 dex for the $1.0 M_{\odot}$). This difference remains until the RGB bump as each model depletes Li along SGB and RGB phases with the same efficiency. \\ Figure~\ref{fig:LiTeffproc} can be compared to Fig.~10 in \citet[][]{2003A&A...399..603P}, where the authors compared the model predictions from classical and rotational models and obtained similar results. We also confirm  that the Li-depletion predicted by Classical models starts at a low effective temperature of $\approx$ 5'700 K, leading to high Li abundances compared to observations and models including rotation. We stress again here the necessity to include rotation to predict consistent Li abundances of low-mass stars. The additional transports of chemicals that we consider for $_{\nu}R1$ model, and $_{\nu(t)}R1$ model enforce the Li-depletion compared to a rotation model without additional mixing, especially for the lowest-mass stars. It should also be taken into account as the amount of Li along the SGB and beginning of RGB is strongly dependent on the one reached at the TAMS. \\

\begin{figure}[ht]
         \center
         \includegraphics [width=90mm]{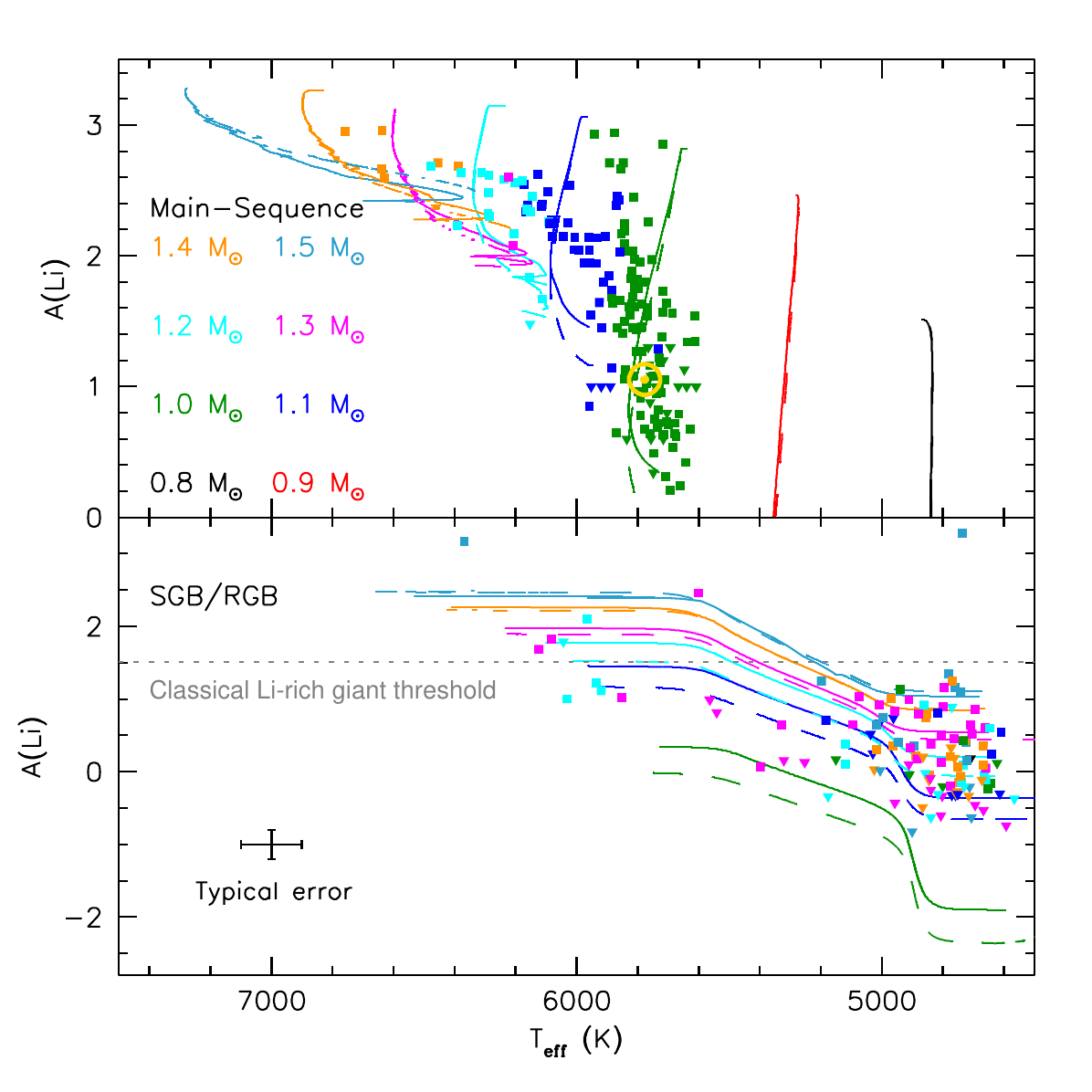}
         \caption{Evolution of Li surface abundance as a function of effective temperature for models at different masses ($0.8-1.5 M_{\odot}$, colour-coded) and solar metallicity. Models including a constant viscosity ($\nu_{\rm add} = 3.5\times10^4$ $\rm cm^2.s^{-1}$) and a time-dependent viscosity (as defined in Sect.~\ref{sect:nuaddt}) are represented in full line and dashed-line, respectively. Top panel: Main sequence. Bottom panel: SGB and RGB phases. The standard Li-rich giant threshold of A(Li) = 1.5 dex is indicated by the grey dashed-line. Observational data are from \citet{2007AJ....133.2464L,2012A&A...541A.150P,2020MNRAS.492..245C,2020A&A...643A.164S,2023ApJ...944L...5M} and represented by colour-coded squares. The Sun is represented in yellow with its usual symbol \citep[][]{2021A&A...653A.141A}.}
         \label{fig:LiTeff}
 \end{figure}

\begin{figure}[h]
         \center
         \includegraphics [width=90mm]{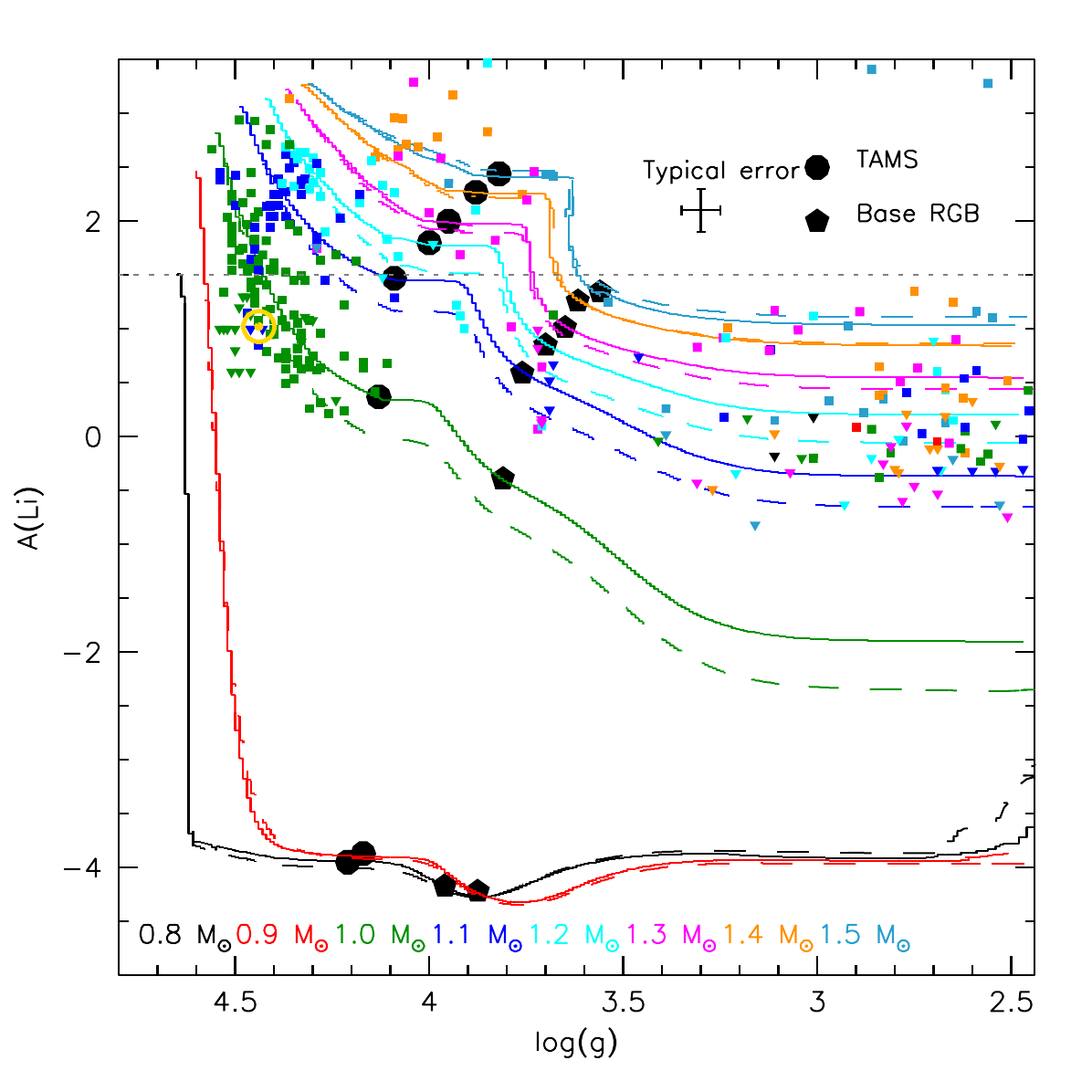}
         \caption{Evolution of Li surface abundance as a function of surface gravity for the same models as in Fig.~\ref{fig:LiTeff}. The standard Li-rich giant threshold of A(Li) = 1.5 dex is indicated by the grey dashed-line. Observational data are from \citet{2007AJ....133.2464L,2012A&A...541A.150P,2020MNRAS.492..245C,2020A&A...643A.164S,2023ApJ...944L...5M} and represented by colour-coded squares. The Sun is represented in yellow with its usual symbol \citep[][]{2021A&A...653A.141A}. TAMS and base of RGB are indicated for $_{\nu}R1$ models by black circles and diamonds, respectively.
         }
         \label{fig:Lievol}
 \end{figure}

Figures \ref{fig:LiTeff} and \ref{fig:Lievol} show the surface Li evolution as a function of effective temperature, and as a function of surface gravity, respectively, with masses from 0.8$M_{\odot}$ to 1.5$M_{\odot}$. Only models $_{\nu}R1$ and $_{\nu(t)}R1$ are represented in these figures.  
As observed for the $1.0 M_{\odot}$ and $1.2 M_{\odot}$ stars, predictions from the updated model result in a stronger Li depletion, mainly along the MS as a consequence of a smaller viscosity at this phase (see Table~\ref{tab:dataLi}). The difference is especially visible for the 1.0, 1.1, and 1.2$M_{\odot}$ stars, for which the angular momentum extraction is high at solar metallicity. The situation is slightly different for the less massive and more massive stars. We recall that the constant value for $\nu_{\rm add}$, as well as the additional chemical transports, was adjusted for the case of the Sun and solar-type stars. It results in a strong Li-depletion for the 0.8 and 0.9 $M_{\odot}$, with almost no Li reaching the SGB with or without the new formalism for the extra viscosity. Predictions for stars more massive than 1.3 $M_{\odot}$ are close in each case all along the evolution as a result of the weak effect of the additional viscosity for such stars that exhibit a vanishing convective envelope.

In the same figures (Figs.~\ref{fig:LiTeff} and \ref{fig:Lievol}), observational constraints from two open clusters (NGC 2420 and M 67) and of the Sun are represented. Additional observational constraints from \citet[][]{2007AJ....133.2464L} and \citet{2023ApJ...944L...5M} are also indicated for giant phases. Each star is colour-coded depending on the corresponding mass determined thanks to our rotation models or from seismic determinations. The uncertainty is small along MS but there is a typical error of 0.2 $M_{\odot}$ for giant stars. The models provide a good fit to Li observations along the MS for stars of $1.0M_{\odot}$ or more massive stars confirming the results obtained in Paper I and II, but a few discrepancies remain along SGB/RGB phases. For the massive stars of our sample, this can mainly be explained by an uncertainty on the determined mass (which is model-dependent) and the different initial rotation velocities of these stars. However, this cannot explain the large difference between the predicted amount of Li and the observed one for the lowest mass stars ($\leq 1.0M_{\odot}$). On the one hand, predictions for the 0.8 and 0.9 $M_{\odot}$ are strongly Li-depleted along the MS in contradiction with observational data that show a depletion of the order of the solar mass star when reaching the RGB phase. The parametric turbulence is not relevant for these stars and should be decreased along the MS. On the other hand, the amount of Li predicted and observed for the 1.0 $M_{\odot}$ is in good agreement with observations along the MS but not anymore when reaching the RGB phase. From the MS to RGB bump ($\approx logg = 2.9$), almost no Li-depletion is observed for these stars whereas models predict a depletion of ~2.3 dex due to the first dredge-up dilution. The discrepancy is mainly due to the fact that we included additional chemical transport to reproduce the Li abundances of solar-type stars in the MS resulting of a small amount of Li when reaching the TAMS ($\approx$ 0.5 dex). Beyond the TAMS, rotation-induced mixing results of a large Li-free region \citep[][]{2003A&A...399..603P,2020A&A...633A..34C} and a strong Li-depletion that is not observed. Consequently, current models cannot explain both the observed MS Li abundance and the RGB Li abundance and call for a process decreasing the Li-depletion along giant phases. However, it has to be mentioned that the RGB stars identified at a stellar mass of 1.0$M_{\odot}$ or less could be more metal-poor than measured. Additional data would be then required to confirm this discrepancy.\\

The significant Li-depletion and potential discrepancy observed for the 1.0M$_{\odot}$ brings an important constraint on the processes involved along the giant phases of evolution. The effect of varying the efficiency of the transport of angular momentum plays a role as well in the chemical evolution. During the MS, due to a strong efficiency for the transport of the angular momentum, additional transport processes were required to reproduce the MS Li-depletion. The involved processes at this phase should then transport efficiently both angular momentum and chemicals before to vanish during the SGB phase. When reaching the sharp transition described in Sect.~\ref{sect:observations}, another process should then transport the angular momentum efficiently but not the chemicals in order to reproduce both observational constraints. This observation supports that two distinct processes are required on each side of the sharp transition. The additional turbulence we considered along MS is confirmed not to be relevant anymore beyond the core rotation rate transition and should be adjusted for the two lowest mass stars of our grid. The penetrative convection should be adapted to the structure of RGB stars, but we dix not expect any additional transport of chemicals to be significant regarding the observations.

\subsection{Li-rich giants threshold}
The correct prediction of the surface Li abundance is especially important for the definition of the so-called Li-rich giant stars, usually defined (from classical model predictions; see also the dotted lines in Fig.~\ref{fig:LiTeffproc}) when A(Li) > 1.5 dex from an initial cosmic abundance of 3.3 dex \citep[e.g.][]{2020A&A...633A..34C}. This threshold is indicated in grey dashed-line in Fig.~\ref{fig:LiTeff}. With respect to a dependence on the mass, \citet[][]{2021MNRAS.505..642D} indicate similar values of A(Li) > 1.9 for a 1.0$M_{\odot}$ and A(Li) > 1.6 for 1.6$M_{\odot}$ (in their introduction). Stars over this limit when reaching the RGB bump are defined as Li-rich giant stars and involve a process of Li enhancement. From that definition, only a few percents of the observed stars are concerned. On the other hand, our results show that we may redefine this threshold from non-standard models and as a function of mass and metallicity as already proposed by \citet{2020A&A...633A..34C}. Our results support smaller thresholds than the classical A(Li) > 1.5 dex (see Table~\ref{tab:dataLi}) and give an estimation of the mass dependence. It is expected as the classical models do not reproduce the correct amount of Li in young open clusters and for the Sun (see e.g. Paper I). We notice however that the effects of the initial rotation velocity, as well as the correct determination of the stellar masses may impact our results if not the general observation (see also Sect.~\ref{sect:CONCLUSION}). The values for the 0.8, 0.9, and 1.0 $M_{\odot}$ have to be taken carefully as well as the Li evolution from TAMS to the RGB bump is uncertain for these low-mass stars as showed in Sect~\ref{sect:Lirgb}. Additional data for the lowest mass stars as well as a dedicated chemical transport analysis along evolution would be required to confirm and specify these results. For the most massive stars of our sample, our results are more robust and show that the classical threshold is overestimated, as already observed by \citet[][]{2020NatAs...4.1059K}. The number of Li-rich giant stars may be higher than commonly found \citep[see also][]{2020A&A...633A..34C}.
 \begin{table}[t]
    \centering
    \caption{Predictions of $_{\nu(t)}R1$ model at different masses for surface Li abundances at the TAMS and RGB bump (columns 2 and 3).}
    \begin{tabular}{c|c|c|c}
         Mass ($M_{\odot}$) & A(Li) TAMS & A(Li) Bump RGB \\
         \hline \hline
         0.8 & -3.95 & -3.80 \\
         0.9 & -3.90 & -3.95 \\
         1.0 & 0.00 & -2.35 \\
         1.1 & 1.20 & -0.70 \\
         1.2 & 1.55 & -0.05 \\
         1.3 & 1.95 & 0.45 \\
         1.4 & 2.25 & 0.90 \\
         1.5 & 2.50 & 1.15 \\
         \hline
    \end{tabular}
    \label{tab:dataLi}
\end{table}

\subsection{Beryllium}
\label{sect:Be}
\begin{figure}[h]
         \center
         \includegraphics [width=90mm]{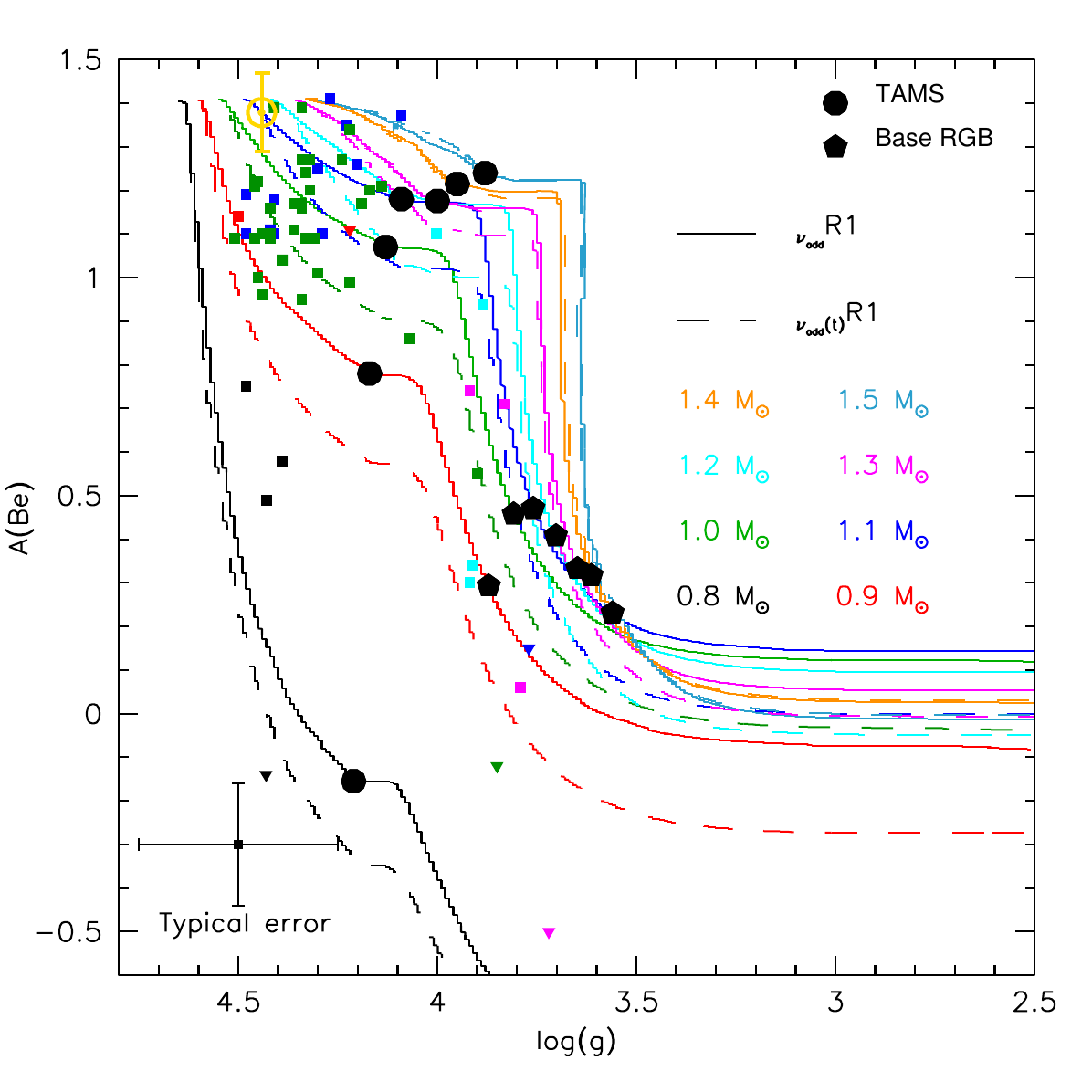}
         \caption{Evolution of Be surface abundance as function of gravity for 0.8 to 1.5 $M_{\odot}$ models. Optimal model predictions are in full line, model without additional transport processes are in dashed-line. Observational data are from \citet[][]{2004A&A...425.1013S,2011A&A...530A..66G,2012ApJ...746...47D,2020ApJ...888...28B,2022ApJ...941...21B}. The Sun is represented in yellow with its usual symbol \citep[][]{2021A&A...653A.141A}.}
         \label{fig:Beevol}
 \end{figure}

As shown for instance by \citet{2020ApJ...888...28B}, an interesting additional constraint in complement to Li is the surface Be abundance evolution. The observation of surface Be in stars is challenging and there is limited data available, Be is destroyed deeper in the stars, at a higher temperature ($\approx$ 3.5 MK) than Li ($\approx$ 2.5 MK) and is expected to be less depleted than Li at a same age. During the PMS and the MS, Be is observed to be only slightly depleted at solar metallicity \citep[see for instance][Paper I]{2003ApJ...582..410B,2003ApJ...583..955B,2004ApJ...605..864B,2022ApJ...941...21B}. \\
Figure~\ref{fig:Beevol} shows the Be surface abundance as a function of gravity for the same models as for Li in Fig.~\ref{fig:Lievol}. Observations of Be are from nine SGB stars from \citet[][]{2020ApJ...888...28B} for the M 67 open cluster and from a compilation of field stars given by \citet[][]{2004A&A...425.1013S,2011A&A...530A..66G,2012ApJ...746...47D,2022ApJ...941...21B}. Models including the modified additional viscosity result for each mass $\le 1.3 M_{\odot}$ of a stronger Be-depletion, as for Li, with a maximum for the 1.0$M_{\odot}$ of -0.15 dex. The depletion is on the same level for the $1.4-1.5 M_{\odot}$ stars. \\ As for the case of Li, a good agreement is obtained between observations and models predictions along the MS. On the other hand, the observed Be-depletion appears to be stronger and to take place earlier than predicted in the SGB phase, and especially when reaching the base of the RGB phase (even considering the error and mass uncertainty). The interpretation of this behaviour is very challenging as it goes in contradiction with the results obtained for Li in Fig.~\ref{fig:Lievol}, where observational constraints support an equal or weaker depletion than the predicted one along the SGB and RGB phases. Also, we used as an initial Be abundance the same value of A(Be) = 1.41 as used by \citet[][]{2020ApJ...888...28B}. A more recent value of A(Be) = 1.30 has been obtained by \citet[][]{2009ARA&A..47..481A}, and would improve the agreement with our results, but the corresponding decrease of 0.1 dex is not sufficient to reconcile observations with our predictions. Additional data for evolved low-mass stars would be required to conclude on Be evolution, but this analysis supports a more efficient transport of Be between the end of the SGB phase and the beginning of the RGB phase in contradiction to Li. 

\section{Summary and discussion} 
\label{sect:CONCLUSION}
We computed models of low-mass stars including additional transport processes for the transport of angular momentum and for chemicals as previously used for PMS/MS stars in Papers I and II. Thanks to the asteroseismic analysis of evolved stars, we have access to the internal rotation of an important sample of SGB and RGB stars that constrain the evolution of the core and surface rotation rates. Exploring the predictions for the models developed in Paper I, and including a constant additional viscosity, we confirmed the issue of a constant transport efficiency for the transport of angular momentum as observed from the MS to the SGB and the RGB phases, and described Sect.\ref{sect:adjust} and for instance by \citet[][]{2019A&A...621A..66E}. This behaviour is consistent with the effect of one or more possibly of two processes that would be involved at two different phases of evolution. Overall, IGW have been shown to efficiently drive the transport during the MS and beginning of the SGB before to vanish reaching the base of the RGB \citep[][]{2005Sci...309.2189C,2008A&A...482..597T,2014MNRAS.444..102K,2017A&A...605A..31P}. The observed transport is also consistent with the effect of magnetohydrodynamic instabilities such as the Tayler-Spruit instability \citep[][]{2002A&A...381..923S,2019MNRAS.485.3661F} or the AMRI \citep[][]{2007MNRAS.377.1481R,2015A&A...573A..80R}. They have been shown to be relevant along the MS and SGB phases as demonstrated recently by \citet{2022NatAs.tmp..119E} for the Sun, and they can also drive transport when reaching the base of the RGB phase. Nevertheless, it remains difficult to conclude on the exact nature of the involved process(es) all along evolution. \\ To simulate the variation over time of the transport of angular momentum, we explored a time-dependent additional viscosity, as described by \citet[][]{2016A&A...589A..23S}, already tested for the solar case in Paper I, and successfully tested by \citet{2023arXiv230207811M} more recently for evolved stars. We modified and re-adjusted the free parameters and confirmed the results obtained by \citet{2023arXiv230207811M} with our evolution code using different prescriptions for the turbulence shear and including a magnetic braking. We showed that such a formalism allows us to aptly reproduce  the behaviour of core and surface rotation rates along course of evolution; first, with a dependence on the differential rotation that allows an almost uniform evolution of the core rotation rates along the RGB phase, while keeping a decreasing core rotation rate along SGB; and, secondly, with a transport efficiency increasing with the mass as observed in previous studies. A first direct update of our formalism would be to add a direct dependence on the metallicity in addition to the mass or, otherwise, directly to the effective temperature. However, such a formalism cannot reproduce the localise sharp transition that takes place while in the SGB phase. 

Nevertheless, this formalism allows us to obtain the correct structure corresponding to the general rotational evolution profile of SGB and RGB stars. Building on our updated model, in agreement with previous studies \citep[e.g.][]{2020A&A...641A.117D,2023arXiv230207811M}, we went to the next step and explored the consequences for the predictions of the surface Li and Be evolution. We compared predictions from a classical model, a rotational model, and rotational models including additional transport processes ($_{\nu}R1$ and $_{\nu(t)}R1$). As already shown by \citet[][]{2006A&A...453..261P}, classical models are not relevant to study surface Li evolution predicting an overly weak depletion along PMS/MS that impacts the amount of Li reaching the TAMS (especially for the lowest mass stars). The updated models, $_{\nu(t)}R1$, including the time-mass dependent viscosity lead to an even stronger depletion during the MS phase, compared to a model with a constant extra viscosity, resulting in smaller abundances from the TAMS to the RGB bump for 1.0 to 1.3 $M_{\odot}$. The new model predictions show that the Li-depletion is strongly dependent on the mass all along evolution. We determined new predictions for Li for the mass-range considered at solar metallicity (see Table~\ref{tab:dataLi}). Our model reproduces the Li-depletion from the MS to the SGB phase well, thanks to the additional transport processes involved along PMS/MS phases (described in Paper I). However, it excepts the 0.8 and 0.9 $M_{\odot}$ that are abnormally depleted and would require a new adjustment of the chemical transport along the MS. \\ When reaching the base of the RGB phase, model predictions are in good agreement with observations for the most massive stars (> 1.1 M$_{\odot}$) and no other chemical transport than first dredge-up dilution and rotational mixing is expected. It highlights that the process that takes place after the sharp transition of the core rotation rate should not drive additional transport of chemicals. The additional turbulence we considered for the MS is not relevant anymore when reaching this transition, nor is a strong transport by penetrative convection. The uncertainty on the mass determination as well as the initial rotation velocity can explain the spread of the observed abundances for these stars. However, it is not the case for the 1.0$M_{\odot}$ for which we predict a weak amount of Li at the RGB bump, in contradiction to the observations (-2.2 dex in place of $\approx$ 0 dex). In other words, no Li-depletion is observed between the TAMS and the RGB bump for this mass. Additional data, with an accurate and robust determination of the metallicity for the lowest mass stars, would be required to confirm this observation but it seems difficult to reconcile MS Li abundance with RGB-bump Li abundance with the present data.  

Our results on Li abundance evolution are also useful in the context of the so-called Li-rich giant stars \citep[hereafter Li-rich giants,][]{2016MNRAS.461.3336C,2019ApJ...880..125C,2020A&A...633A..34C,2021A&A...651A..84M,2021NatAs...5...86Y,2021MNRAS.505..642D}. They are commonly defined when a value of A(Li) $>$ 1.5 dex is observed at evolved phase (from a cosmological value of A(Li) = 3.3 dex). This threshold corresponds to the mean value predicted by a classical stellar evolution models (including only transport by convection) after the first dredge-up. Classical models predict almost no surface Li for RGB stars \citep[][and see Fig.~\ref{fig:LiTeffproc}]{2012A&A...543A.108L}, mainly due to a severe depletion during the first dredge-up when Li is destroyed in the deep hot layers of the star. However, the threshold of A(Li) $>$ 1.5 dex has to be taken carefully as it usually does not take into account the constraints on the chemicals along the MS. An unknown process is invoked in order to explain the enhancement of Li observed at this phase for a few stars. We stress out that our results support that this threshold is overestimated and should be dependent on the mass and the metallicity as already observed by \citet[][]{2016ApJ...819..135K,2020A&A...633A..34C,2020A&A...635A.142K}. The robust determination of the evolutionary phase (RGB or red clump stars) of the Li-rich giant stars is an issue as well \citep[see e.g. ][]{2021NatAs...5...86Y}, as it may involve different transport processes. It also impacts the definition of the threshold, as it is not expected to be the same depending on the phase \citep[e.g.][who proposed a new threshold of A(Li) = - 0.9 dex for red clump stars]{2020NatAs...4.1059K}.
A better discrimination among the Li-rich stars and "Li-normal" stars at different masses and different metallicities is a step forward to the identification of the process at the origin of the Li-enhancement. A key to bring important additional constraints would be the availability of asteroseismic targets for which Li abundances would be determined. On the other hand, the amount of Li expected along stellar evolution is highly dependent on the physics included in the stellar models from the PMS to the evolved phases. We highlight the importance of reproducing the MS Li abundances to get the correct SGB/RGB Li abundance, especially for the solar-mass stars. Then, the combination with predictions from non-standard stellar models is necessary in order to obtain a correct understanding. 

Observational constraints on Be also bring additional information on the nature of the processes involved along the SGB/RGB phases. Our results show that Be-depletion is correctly predicted by our models, compared to the observations along the MS and SGB phases. This stands in contradiction to the results on Li, but we need additional observations to conclude on Be, especially along the RGB phase.

\begin{acknowledgements}
This work was supported by the European Union (ChETEC-INFRA, project no. 101008324). The author thanks A. Bonhomme, M. Pinsonneault, and M. Cantiello for helpful discussions, and the anonymous referee for constructive comments on the manuscript. Special thanks go to J. Kerutt for discussions all along this work. The author also thanks the Department of Astronomy of the University of Geneva for hosting and support during part of this work. This research has made use of NASA's Astrophysics Data System Bibliographic Services.  
\end{acknowledgements}

%
%

\bibliographystyle{aa}
\bibliography{references}

\begin{thebibliography}{111}
\expandafter\ifx\csname natexlab\endcsname\relax\def\natexlab#1{#1}\fi

\bibitem[{{Aerts} {et~al.}(2021){Aerts}, {Augustson}, {Mathis}, {Pedersen},
  {Mombarg}, {Vanlaer}, {Van Beeck}, \& {Van Reeth}}]{2021A&A...656A.121A}
{Aerts}, C., {Augustson}, K., {Mathis}, S., {et~al.} 2021, \aap, 656, A121

\bibitem[{{Aerts} {et~al.}(2019){Aerts}, {Mathis}, \&
  {Rogers}}]{2019ARA&A..57...35A}
{Aerts}, C., {Mathis}, S., \& {Rogers}, T.~M. 2019, \araa, 57, 35

\bibitem[{{Amard} {et~al.}(2016){Amard}, {Palacios}, {Charbonnel}, {Gallet}, \&
  {Bouvier}}]{2016A&A...587A.105A}
{Amard}, L., {Palacios}, A., {Charbonnel}, C., {Gallet}, F., \& {Bouvier}, J.
  2016, \aap, 587, A105

\bibitem[{{Amard} {et~al.}(2019){Amard}, {Palacios}, {Charbonnel}, {Gallet},
  {Georgy}, {Lagarde}, \& {Siess}}]{2019A&A...631A..77A}
{Amard}, L., {Palacios}, A., {Charbonnel}, C., {et~al.} 2019, \aap, 631, A77

\bibitem[{{Anders} \& {Grevesse}(1989)}]{1989GeCoA..53..197A}
{Anders}, E. \& {Grevesse}, N. 1989, \gca, 53, 197

\bibitem[{{Asplund} {et~al.}(2021){Asplund}, {Amarsi}, \&
  {Grevesse}}]{2021A&A...653A.141A}
{Asplund}, M., {Amarsi}, A.~M., \& {Grevesse}, N. 2021, \aap, 653, A141

\bibitem[{{Asplund} {et~al.}(2009){Asplund}, {Grevesse}, {Sauval}, \&
  {Scott}}]{2009ARA&A..47..481A}
{Asplund}, M., {Grevesse}, N., {Sauval}, A.~J., \& {Scott}, P. 2009, \araa, 47,
  481

\bibitem[{{Augustson} \& {Mathis}(2019)}]{2019ApJ...874...83A}
{Augustson}, K.~C. \& {Mathis}, S. 2019, \apj, 874, 83

\bibitem[{{Balachandran}(1995)}]{1995ApJ...446..203B}
{Balachandran}, S. 1995, \apj, 446, 203

\bibitem[{{Belkacem} {et~al.}(2015){Belkacem}, {Marques}, {Goupil}, {Mosser},
  {Sonoi}, {Ouazzani}, {Dupret}, {Mathis}, \& {Grosjean}}]{2015A&A...579A..31B}
{Belkacem}, K., {Marques}, J.~P., {Goupil}, M.~J., {et~al.} 2015, \aap, 579,
  A31

\bibitem[{{Benomar} {et~al.}(2015){Benomar}, {Takata}, {Shibahashi},
  {Ceillier}, \& {Garc{\'\i}a}}]{2015MNRAS.452.2654B}
{Benomar}, O., {Takata}, M., {Shibahashi}, H., {Ceillier}, T., \&
  {Garc{\'\i}a}, R.~A. 2015, \mnras, 452, 2654

\bibitem[{{Boesgaard}(1976)}]{1976PASP...88..353B}
{Boesgaard}, A.~M. 1976, \pasp, 88, 353

\bibitem[{{Boesgaard} {et~al.}(2003{\natexlab{a}}){Boesgaard}, {Armengaud}, \&
  {King}}]{2003ApJ...583..955B}
{Boesgaard}, A.~M., {Armengaud}, E., \& {King}, J.~R. 2003{\natexlab{a}}, \apj,
  583, 955

\bibitem[{{Boesgaard} {et~al.}(2003{\natexlab{b}}){Boesgaard}, {Armengaud}, \&
  {King}}]{2003ApJ...582..410B}
{Boesgaard}, A.~M., {Armengaud}, E., \& {King}, J.~R. 2003{\natexlab{b}}, \apj,
  582, 410

\bibitem[{{Boesgaard} {et~al.}(2004){Boesgaard}, {Armengaud}, \&
  {King}}]{2004ApJ...605..864B}
{Boesgaard}, A.~M., {Armengaud}, E., \& {King}, J.~R. 2004, \apj, 605, 864

\bibitem[{{Boesgaard} {et~al.}(2022{\natexlab{a}}){Boesgaard}, {Deliyannis},
  {Lum}, \& {Chontos}}]{2022ApJ...941...21B}
{Boesgaard}, A.~M., {Deliyannis}, C.~P., {Lum}, M.~G., \& {Chontos}, A.
  2022{\natexlab{a}}, \apj, 941, 21

\bibitem[{{Boesgaard} {et~al.}(2022{\natexlab{b}}){Boesgaard}, {Lum},
  {Chontos}, \& {Deliyannis}}]{2022ApJ...927..118B}
{Boesgaard}, A.~M., {Lum}, M.~G., {Chontos}, A., \& {Deliyannis}, C.~P.
  2022{\natexlab{b}}, \apj, 927, 118

\bibitem[{{Boesgaard} {et~al.}(2020){Boesgaard}, {Lum}, \&
  {Deliyannis}}]{2020ApJ...888...28B}
{Boesgaard}, A.~M., {Lum}, M.~G., \& {Deliyannis}, C.~P. 2020, \apj, 888, 28

\bibitem[{{Boesgaard} {et~al.}(2016){Boesgaard}, {Lum}, {Deliyannis}, {King},
  {Pinsonneault}, \& {Somers}}]{2016ApJ...830...49B}
{Boesgaard}, A.~M., {Lum}, M.~G., {Deliyannis}, C.~P., {et~al.} 2016, \apj,
  830, 49

\bibitem[{{Boesgaard} \& {Tripicco}(1986)}]{1986ApJ...302L..49B}
{Boesgaard}, A.~M. \& {Tripicco}, M.~J. 1986, \apjl, 302, L49

\bibitem[{{Bossini} {et~al.}(2019){Bossini}, {Vallenari}, {Bragaglia},
  {Cantat-Gaudin}, {Sordo}, {Balaguer-N{\'u}{\~n}ez}, {Jordi}, {Moitinho},
  {Soubiran}, {Casamiquela}, {Carrera}, \& {Heiter}}]{2019A&A...623A.108B}
{Bossini}, D., {Vallenari}, A., {Bragaglia}, A., {et~al.} 2019, \aap, 623, A108

\bibitem[{{Carlos} {et~al.}(2020){Carlos}, {Mel{\'e}ndez}, {do Nascimento}, \&
  {Castro}}]{2020MNRAS.492..245C}
{Carlos}, M., {Mel{\'e}ndez}, J., {do Nascimento}, J.-D., \& {Castro}, M. 2020,
  \mnras, 492, 245

\bibitem[{{Carlos} {et~al.}(2019){Carlos}, {Mel{\'e}ndez}, {Spina}, {dos
  Santos}, {Bedell}, {Ramirez}, {Asplund}, {Bean}, {Yong}, {Yana Galarza}, \&
  {Alves-Brito}}]{2019MNRAS.485.4052C}
{Carlos}, M., {Mel{\'e}ndez}, J., {Spina}, L., {et~al.} 2019, \mnras, 485, 4052

\bibitem[{{Casey} {et~al.}(2019){Casey}, {Ho}, {Ness}, {Hogg}, {Rix},
  {Angelou}, {Hekker}, {Tout}, {Lattanzio}, {Karakas}, {Woods}, {Price-Whelan},
  \& {Schlaufman}}]{2019ApJ...880..125C}
{Casey}, A.~R., {Ho}, A. Y.~Q., {Ness}, M., {et~al.} 2019, \apj, 880, 125

\bibitem[{{Casey} {et~al.}(2016){Casey}, {Ruchti}, {Masseron}, {Randich},
  {Gilmore}, {Lind}, {Kennedy}, {Koposov}, {Hourihane}, {Franciosini}, {Lewis},
  {Magrini}, {Morbidelli}, {Sacco}, {Worley}, {Feltzing}, {Jeffries},
  {Vallenari}, {Bensby}, {Bragaglia}, {Flaccomio}, {Francois}, {Korn},
  {Lanzafame}, {Pancino}, {Recio-Blanco}, {Smiljanic}, {Carraro}, {Costado},
  {Damiani}, {Donati}, {Frasca}, {Jofr{\'e}}, {Lardo}, {de Laverny}, {Monaco},
  {Prisinzano}, {Sbordone}, {Sousa}, {Tautvai{\v{s}}ien{\.{e}}}, {Zaggia},
  {Zwitter}, {Delgado Mena}, {Chorniy}, {Martell}, {Silva Aguirre}, {Miglio},
  {Chiappini}, {Montalban}, {Morel}, \& {Valentini}}]{2016MNRAS.461.3336C}
{Casey}, A.~R., {Ruchti}, G., {Masseron}, T., {et~al.} 2016, \mnras, 461, 3336

\bibitem[{{Ceillier} {et~al.}(2013){Ceillier}, {Eggenberger}, {Garc{\'\i}a}, \&
  {Mathis}}]{2013A&A...555A..54C}
{Ceillier}, T., {Eggenberger}, P., {Garc{\'\i}a}, R.~A., \& {Mathis}, S. 2013,
  \aap, 555, A54

\bibitem[{{Chanam{\'e}} {et~al.}(2022){Chanam{\'e}}, {Pinsonneault},
  {Aguilera-G{\'o}mez}, \& {Zinn}}]{2022ApJ...933...58C}
{Chanam{\'e}}, J., {Pinsonneault}, M.~H., {Aguilera-G{\'o}mez}, C., \& {Zinn},
  J.~C. 2022, \apj, 933, 58

\bibitem[{{Charbonnel} {et~al.}(2021){Charbonnel}, {Borisov}, {de Laverny}, \&
  {Prantzos}}]{2021A&A...649L..10C}
{Charbonnel}, C., {Borisov}, S., {de Laverny}, P., \& {Prantzos}, N. 2021,
  \aap, 649, L10

\bibitem[{{Charbonnel} {et~al.}(2013){Charbonnel}, {Decressin}, {Amard},
  {Palacios}, \& {Talon}}]{2013A&A...554A..40C}
{Charbonnel}, C., {Decressin}, T., {Amard}, L., {Palacios}, A., \& {Talon}, S.
  2013, \aap, 554, A40

\bibitem[{{Charbonnel} {et~al.}(2020){Charbonnel}, {Lagarde}, {Jasniewicz},
  {North}, {Shetrone}, {Krugler Hollek}, {Smith}, {Smiljanic}, {Palacios}, \&
  {Ottoni}}]{2020A&A...633A..34C}
{Charbonnel}, C., {Lagarde}, N., {Jasniewicz}, G., {et~al.} 2020, \aap, 633,
  A34

\bibitem[{{Charbonnel} \& {Talon}(2005)}]{2005Sci...309.2189C}
{Charbonnel}, C. \& {Talon}, S. 2005, Science, 309, 2189

\bibitem[{{Charbonnel} {et~al.}(1992){Charbonnel}, {Vauclair}, \&
  {Zahn}}]{1992A&A...255..191C}
{Charbonnel}, C., {Vauclair}, S., \& {Zahn}, J.~P. 1992, \aap, 255, 191

\bibitem[{{Cummings} {et~al.}(2017){Cummings}, {Deliyannis}, {Maderak}, \&
  {Steinhauer}}]{2017AJ....153..128C}
{Cummings}, J.~D., {Deliyannis}, C.~P., {Maderak}, R.~M., \& {Steinhauer}, A.
  2017, \aj, 153, 128

\bibitem[{{Decressin} {et~al.}(2009){Decressin}, {Mathis}, {Palacios}, {Siess},
  {Talon}, {Charbonnel}, \& {Zahn}}]{2009A&A...495..271D}
{Decressin}, T., {Mathis}, S., {Palacios}, A., {et~al.} 2009, \aap, 495, 271

\bibitem[{{Deepak} \& {Lambert}(2021)}]{2021MNRAS.505..642D}
{Deepak} \& {Lambert}, D.~L. 2021, \mnras, 505, 642

\bibitem[{{Deheuvels} {et~al.}(2015){Deheuvels}, {Ballot}, {Beck}, {Mosser},
  {{\O}stensen}, {Garc{\'\i}a}, \& {Goupil}}]{2015A&A...580A..96D}
{Deheuvels}, S., {Ballot}, J., {Beck}, P.~G., {et~al.} 2015, \aap, 580, A96

\bibitem[{{Deheuvels} {et~al.}(2020){Deheuvels}, {Ballot}, {Eggenberger},
  {Spada}, {Noll}, \& {den Hartogh}}]{2020A&A...641A.117D}
{Deheuvels}, S., {Ballot}, J., {Eggenberger}, P., {et~al.} 2020, \aap, 641,
  A117

\bibitem[{{Deheuvels} {et~al.}(2014){Deheuvels}, {Do{\u{g}}an}, {Goupil},
  {Appourchaux}, {Benomar}, {Bruntt}, {Campante}, {Casagrande}, {Ceillier},
  {Davies}, {De Cat}, {Fu}, {Garc{\'\i}a}, {Lobel}, {Mosser}, {Reese},
  {Regulo}, {Schou}, {Stahn}, {Thygesen}, {Yang}, {Chaplin},
  {Christensen-Dalsgaard}, {Eggenberger}, {Gizon}, {Mathis},
  {Molenda-{\.Z}akowicz}, \& {Pinsonneault}}]{2014A&A...564A..27D}
{Deheuvels}, S., {Do{\u{g}}an}, G., {Goupil}, M.~J., {et~al.} 2014, \aap, 564,
  A27

\bibitem[{{Deheuvels} {et~al.}(2012){Deheuvels}, {Garc{\'\i}a}, {Chaplin},
  {Basu}, {Antia}, {Appourchaux}, {Benomar}, {Davies}, {Elsworth}, {Gizon},
  {Goupil}, {Reese}, {Regulo}, {Schou}, {Stahn}, {Casagrande},
  {Christensen-Dalsgaard}, {Fischer}, {Hekker}, {Kjeldsen}, {Mathur}, {Mosser},
  {Pinsonneault}, {Valenti}, {Christiansen}, {Kinemuchi}, \&
  {Mullally}}]{2012ApJ...756...19D}
{Deheuvels}, S., {Garc{\'\i}a}, R.~A., {Chaplin}, W.~J., {et~al.} 2012, \apj,
  756, 19

\bibitem[{{Delgado Mena} {et~al.}(2012){Delgado Mena}, {Israelian},
  {Gonz{\'a}lez Hern{\'a}ndez}, {Santos}, \& {Rebolo}}]{2012ApJ...746...47D}
{Delgado Mena}, E., {Israelian}, G., {Gonz{\'a}lez Hern{\'a}ndez}, J.~I.,
  {Santos}, N.~C., \& {Rebolo}, R. 2012, \apj, 746, 47

\bibitem[{{Deliyannis} {et~al.}(2019){Deliyannis}, {Anthony-Twarog},
  {Lee-Brown}, \& {Twarog}}]{2019AJ....158..163D}
{Deliyannis}, C.~P., {Anthony-Twarog}, B.~J., {Lee-Brown}, D.~B., \& {Twarog},
  B.~A. 2019, \aj, 158, 163

\bibitem[{{Deliyannis} {et~al.}(2000){Deliyannis}, {Pinsonneault}, \&
  {Charbonnel}}]{2000IAUS..198...61D}
{Deliyannis}, C.~P., {Pinsonneault}, M.~H., \& {Charbonnel}, C. 2000, in IAU
  Symposium, Vol. 198, The Light Elements and their Evolution, ed. L.~{da
  Silva}, R.~{de Medeiros}, \& M.~{Spite}, 61

\bibitem[{{Denissenkov} {et~al.}(2010){Denissenkov}, {Pinsonneault},
  {Terndrup}, \& {Newsham}}]{2010ApJ...716.1269D}
{Denissenkov}, P.~A., {Pinsonneault}, M., {Terndrup}, D.~M., \& {Newsham}, G.
  2010, \apj, 716, 1269

\bibitem[{{Di Mauro} {et~al.}(2016){Di Mauro}, {Ventura}, {Cardini}, {Stello},
  {Christensen-Dalsgaard}, {Dziembowski}, {Patern{\`o}}, {Beck}, {Bloemen},
  {Davies}, {De Smedt}, {Elsworth}, {Garc{\'\i}a}, {Hekker}, {Mosser}, \&
  {Tkachenko}}]{2016ApJ...817...65D}
{Di Mauro}, M.~P., {Ventura}, R., {Cardini}, D., {et~al.} 2016, \apj, 817, 65

\bibitem[{{Do Nascimento} {et~al.}(2009){Do Nascimento}, {Castro},
  {Mel{\'e}ndez}, {Bazot}, {Th{\'e}ado}, {Porto de Mello}, \& {de
  Medeiros}}]{2009A&A...501..687D}
{Do Nascimento}, J.~D., J., {Castro}, M., {Mel{\'e}ndez}, J., {et~al.} 2009,
  \aap, 501, 687

\bibitem[{{Dumont} {et~al.}(2021{\natexlab{a}}){Dumont}, {Charbonnel},
  {Palacios}, \& {Borisov}}]{2021A&A...654A..46D}
{Dumont}, T., {Charbonnel}, C., {Palacios}, A., \& {Borisov}, S.
  2021{\natexlab{a}}, \aap, 654, A46

\bibitem[{{Dumont} {et~al.}(2021{\natexlab{b}}){Dumont}, {Palacios},
  {Charbonnel}, {Richard}, {Amard}, {Augustson}, \&
  {Mathis}}]{2021A&A...646A..48D}
{Dumont}, T., {Palacios}, A., {Charbonnel}, C., {et~al.} 2021{\natexlab{b}},
  \aap, 646, A48

\bibitem[{{Eggenberger} {et~al.}(2019{\natexlab{a}}){Eggenberger}, {Buldgen},
  \& {Salmon}}]{2019A&A...626L...1E}
{Eggenberger}, P., {Buldgen}, G., \& {Salmon}, S.~J.~A.~J. 2019{\natexlab{a}},
  \aap, 626, L1

\bibitem[{{Eggenberger} {et~al.}(2022){Eggenberger}, {Buldgen}, {Salmon},
  {Noels}, {Grevesse}, \& {Asplund}}]{2022NatAs.tmp..119E}
{Eggenberger}, P., {Buldgen}, G., {Salmon}, S.~J.~A.~J., {et~al.} 2022, Nature
  Astronomy

\bibitem[{{Eggenberger} {et~al.}(2019{\natexlab{b}}){Eggenberger}, {Deheuvels},
  {Miglio}, {Ekstr{\"o}m}, {Georgy}, {Meynet}, {Lagarde}, {Salmon}, {Buldgen},
  {Montalb{\'a}n}, {Spada}, \& {Ballot}}]{2019A&A...621A..66E}
{Eggenberger}, P., {Deheuvels}, S., {Miglio}, A., {et~al.} 2019{\natexlab{b}},
  Astronomy and Astrophysics, 621, A66

\bibitem[{{Eggenberger} {et~al.}(2019{\natexlab{c}}){Eggenberger}, {den
  Hartogh}, {Buldgen}, {Meynet}, {Salmon}, \&
  {Deheuvels}}]{2019A&A...631L...6E}
{Eggenberger}, P., {den Hartogh}, J.~W., {Buldgen}, G., {et~al.}
  2019{\natexlab{c}}, \aap, 631, L6

\bibitem[{{Eggenberger} {et~al.}(2017){Eggenberger}, {Lagarde}, {Miglio},
  {Montalb{\'a}n}, {Ekstr{\"o}m}, {Georgy}, {Meynet}, {Salmon}, {Ceillier},
  {Garc{\'\i}a}, {Mathis}, {Deheuvels}, {Maeder}, {den Hartogh}, \&
  {Hirschi}}]{2017A&A...599A..18E}
{Eggenberger}, P., {Lagarde}, N., {Miglio}, A., {et~al.} 2017, \aap, 599, A18

\bibitem[{{Eggenberger} {et~al.}(2005){Eggenberger}, {Maeder}, \&
  {Meynet}}]{2005A&A...440L...9E}
{Eggenberger}, P., {Maeder}, A., \& {Meynet}, G. 2005, \aap, 440, L9

\bibitem[{{Eggenberger} {et~al.}(2010){Eggenberger}, {Meynet}, {Maeder},
  {Miglio}, {Montalban}, {Carrier}, {Mathis}, {Charbonnel}, \&
  {Talon}}]{2010A&A...519A.116E}
{Eggenberger}, P., {Meynet}, G., {Maeder}, A., {et~al.} 2010, \aap, 519, A116

\bibitem[{{Eggenberger} {et~al.}(2012){Eggenberger}, {Montalb{\'a}n}, \&
  {Miglio}}]{2012A&A...544L...4E}
{Eggenberger}, P., {Montalb{\'a}n}, J., \& {Miglio}, A. 2012, Astronomy and
  Astrophysics, 544, L4

\bibitem[{{Fuller} {et~al.}(2014){Fuller}, {Lecoanet}, {Cantiello}, \&
  {Brown}}]{2014ApJ...796...17F}
{Fuller}, J., {Lecoanet}, D., {Cantiello}, M., \& {Brown}, B. 2014, \apj, 796,
  17

\bibitem[{{Fuller} {et~al.}(2019){Fuller}, {Piro}, \&
  {Jermyn}}]{2019MNRAS.485.3661F}
{Fuller}, J., {Piro}, A.~L., \& {Jermyn}, A.~S. 2019, \mnras, 485, 3661

\bibitem[{{Gallet} \& {Bouvier}(2015)}]{2015A&A...577A..98G}
{Gallet}, F. \& {Bouvier}, J. 2015, \aap, 577, A98

\bibitem[{{G{\'a}lvez-Ortiz} {et~al.}(2011){G{\'a}lvez-Ortiz}, {Delgado-Mena},
  {Gonz{\'a}lez Hern{\'a}ndez}, {Israelian}, {Santos}, {Rebolo}, \&
  {Ecuvillon}}]{2011A&A...530A..66G}
{G{\'a}lvez-Ortiz}, M.~C., {Delgado-Mena}, E., {Gonz{\'a}lez Hern{\'a}ndez},
  J.~I., {et~al.} 2011, \aap, 530, A66

\bibitem[{{Garc{\'\i}a} \& {Ballot}(2019)}]{2019LRSP...16....4G}
{Garc{\'\i}a}, R.~A. \& {Ballot}, J. 2019, Living Reviews in Solar Physics, 16,
  4

\bibitem[{{Gehan} {et~al.}(2018){Gehan}, {Mosser}, {Michel}, {Samadi}, \&
  {Kallinger}}]{2018A&A...616A..24G}
{Gehan}, C., {Mosser}, B., {Michel}, E., {Samadi}, R., \& {Kallinger}, T. 2018,
  \aap, 616, A24

\bibitem[{{Goupil} {et~al.}(2013){Goupil}, {Mosser}, {Marques}, {Ouazzani},
  {Belkacem}, {Lebreton}, \& {Samadi}}]{2013A&A...549A..75G}
{Goupil}, M.~J., {Mosser}, B., {Marques}, J.~P., {et~al.} 2013, \aap, 549, A75

\bibitem[{{Kirby} {et~al.}(2016){Kirby}, {Guhathakurta}, {Zhang}, {Hong},
  {Guo}, {Guo}, {Cohen}, \& {Cunha}}]{2016ApJ...819..135K}
{Kirby}, E.~N., {Guhathakurta}, P., {Zhang}, A.~J., {et~al.} 2016, \apj, 819,
  135

\bibitem[{{Kissin} \& {Thompson}(2015)}]{2015ApJ...808...35K}
{Kissin}, Y. \& {Thompson}, C. 2015, \apj, 808, 35

\bibitem[{{Kjeldsen} \& {Bedding}(1995)}]{1995A&A...293...87K}
{Kjeldsen}, H. \& {Bedding}, T.~R. 1995, \aap, 293, 87

\bibitem[{{Kochukhov} {et~al.}(2020){Kochukhov}, {Hackman}, {Lehtinen}, \&
  {Wehrhahn}}]{2020A&A...635A.142K}
{Kochukhov}, O., {Hackman}, T., {Lehtinen}, J.~J., \& {Wehrhahn}, A. 2020,
  \aap, 635, A142

\bibitem[{{Kumar} {et~al.}(2020){Kumar}, {Reddy}, {Campbell}, {Maben}, {Zhao},
  \& {Ting}}]{2020NatAs...4.1059K}
{Kumar}, Y.~B., {Reddy}, B.~E., {Campbell}, S.~W., {et~al.} 2020, Nature
  Astronomy, 4, 1059

\bibitem[{{Kurtz} {et~al.}(2014){Kurtz}, {Saio}, {Takata}, {Shibahashi},
  {Murphy}, \& {Sekii}}]{2014MNRAS.444..102K}
{Kurtz}, D.~W., {Saio}, H., {Takata}, M., {et~al.} 2014, \mnras, 444, 102

\bibitem[{{Lagarde} {et~al.}(2012){Lagarde}, {Decressin}, {Charbonnel},
  {Eggenberger}, {Ekstr{\"o}m}, \& {Palacios}}]{2012A&A...543A.108L}
{Lagarde}, N., {Decressin}, T., {Charbonnel}, C., {et~al.} 2012, \aap, 543,
  A108

\bibitem[{{Lodders}(2003)}]{2003ApJ...591.1220L}
{Lodders}, K. 2003, \apj, 591, 1220

\bibitem[{{Luck} \& {Heiter}(2007)}]{2007AJ....133.2464L}
{Luck}, R.~E. \& {Heiter}, U. 2007, \aj, 133, 2464

\bibitem[{{Maeder} \& {Zahn}(1998)}]{1998A&A...334.1000M}
{Maeder}, A. \& {Zahn}, J.-P. 1998, \aap, 334, 1000

\bibitem[{{Magrini} {et~al.}(2021){Magrini}, {Lagarde}, {Charbonnel},
  {Franciosini}, {Randich}, {Smiljanic}, {Casali}, {Viscasillas V{\'a}zquez},
  {Spina}, {Biazzo}, {Pasquini}, {Bragaglia}, {Van der Swaelmen},
  {Tautvai{\v{s}}ien{\.{e}}}, {Inno}, {Sanna}, {Prisinzano}, {Degl'Innocenti},
  {Prada Moroni}, {Roccatagliata}, {Tognelli}, {Monaco}, {de Laverny},
  {Delgado-Mena}, {Baratella}, {D'Orazi}, {Vallenari}, {Gonneau}, {Worley},
  {Jim{\'e}nez-Esteban}, {Jofre}, {Bensby}, {Fran{\c{c}}ois}, {Guiglion},
  {Bayo}, {Jeffries}, {Binks}, {Gilmore}, {Damiani}, {Korn}, {Pancino},
  {Sacco}, {Hourihane}, {Morbidelli}, \& {Zaggia}}]{2021A&A...651A..84M}
{Magrini}, L., {Lagarde}, N., {Charbonnel}, C., {et~al.} 2021, \aap, 651, A84

\bibitem[{{Mallick} {et~al.}(2023){Mallick}, {Singh}, \&
  {Reddy}}]{2023ApJ...944L...5M}
{Mallick}, A., {Singh}, R., \& {Reddy}, B.~E. 2023, \apjl, 944, L5

\bibitem[{{Marques} {et~al.}(2013){Marques}, {Goupil}, {Lebreton}, {Talon},
  {Palacios}, {Belkacem}, {Ouazzani}, {Mosser}, {Moya}, {Morel}, {Pichon},
  {Mathis}, {Zahn}, {Turck-Chi{\`e}ze}, \& {Nghiem}}]{2013A&A...549A..74M}
{Marques}, J.~P., {Goupil}, M.~J., {Lebreton}, Y., {et~al.} 2013, \aap, 549,
  A74

\bibitem[{{Mathis} {et~al.}(2018){Mathis}, {Prat}, {Amard}, {Charbonnel},
  {Palacios}, {Lagarde}, \& {Eggenberger}}]{2018A&A...620A..22M}
{Mathis}, S., {Prat}, V., {Amard}, L., {et~al.} 2018, Astronomy and
  Astrophysics, 620, A22

\bibitem[{{Mathis} \& {Zahn}(2004)}]{2004A&A...425..229M}
{Mathis}, S. \& {Zahn}, J.~P. 2004, \aap, 425, 229

\bibitem[{{Mathis} \& {Zahn}(2005)}]{2005A&A...440..653M}
{Mathis}, S. \& {Zahn}, J.~P. 2005, \aap, 440, 653

\bibitem[{{Matt} {et~al.}(2015){Matt}, {Brun}, {Baraffe}, {Bouvier}, \&
  {Chabrier}}]{Matt2015}
{Matt}, S.~P., {Brun}, A.~S., {Baraffe}, I., {Bouvier}, J., \& {Chabrier}, G.
  2015, \apjl, 799, L23

\bibitem[{{Mosser} {et~al.}(2012){Mosser}, {Goupil}, {Belkacem}, {Marques},
  {Beck}, {Bloemen}, {De Ridder}, {Barban}, {Deheuvels}, {Elsworth}, {Hekker},
  {Kallinger}, {Ouazzani}, {Pinsonneault}, {Samadi}, {Stello}, {Garc{\'\i}a},
  {Klaus}, {Li}, {Mathur}, \& {Morris}}]{2012A&A...548A..10M}
{Mosser}, B., {Goupil}, M.~J., {Belkacem}, K., {et~al.} 2012, \aap, 548, A10

\bibitem[{{Mosser} {et~al.}(2013){Mosser}, {Michel}, {Belkacem}, {Goupil},
  {Baglin}, {Barban}, {Provost}, {Samadi}, {Auvergne}, \&
  {Catala}}]{2013A&A...550A.126M}
{Mosser}, B., {Michel}, E., {Belkacem}, K., {et~al.} 2013, \aap, 550, A126

\bibitem[{{Moyano} {et~al.}(2022){Moyano}, {Eggenberger}, {Meynet}, {Gehan},
  {Mosser}, {Buldgen}, \& {Salmon}}]{2022A&A...663A.180M}
{Moyano}, F.~D., {Eggenberger}, P., {Meynet}, G., {et~al.} 2022, \aap, 663,
  A180

\bibitem[{{Moyano} {et~al.}(2023){Moyano}, {Eggenberger}, {Mosser}, \&
  {Spada}}]{2023arXiv230207811M}
{Moyano}, F.~D., {Eggenberger}, P., {Mosser}, B., \& {Spada}, F. 2023, arXiv
  e-prints, arXiv:2302.07811

\bibitem[{{Pace} {et~al.}(2012){Pace}, {Castro}, {Mel{\'e}ndez}, {Th{\'e}ado},
  \& {do Nascimento}}]{2012A&A...541A.150P}
{Pace}, G., {Castro}, M., {Mel{\'e}ndez}, J., {Th{\'e}ado}, S., \& {do
  Nascimento}, J.~D., J. 2012, \aap, 541, A150

\bibitem[{{Palacios} {et~al.}(2006){Palacios}, {Charbonnel}, {Talon}, \&
  {Siess}}]{2006A&A...453..261P}
{Palacios}, A., {Charbonnel}, C., {Talon}, S., \& {Siess}, L. 2006, \aap, 453,
  261

\bibitem[{{Palacios} {et~al.}(2003){Palacios}, {Talon}, {Charbonnel}, \&
  {Forestini}}]{2003A&A...399..603P}
{Palacios}, A., {Talon}, S., {Charbonnel}, C., \& {Forestini}, M. 2003, \aap,
  399, 603

\bibitem[{{Paquette} {et~al.}(1986){Paquette}, {Pelletier}, {Fontaine}, \&
  {Michaud}}]{1986ApJS...61..177P}
{Paquette}, C., {Pelletier}, C., {Fontaine}, G., \& {Michaud}, G. 1986, \apjs,
  61, 177

\bibitem[{{Pin{\c{c}}on} {et~al.}(2017){Pin{\c{c}}on}, {Belkacem}, {Goupil}, \&
  {Marques}}]{2017A&A...605A..31P}
{Pin{\c{c}}on}, C., {Belkacem}, K., {Goupil}, M.~J., \& {Marques}, J.~P. 2017,
  \aap, 605, A31

\bibitem[{{Pinsonneault} {et~al.}(1990){Pinsonneault}, {Kawaler}, \&
  {Demarque}}]{1990ApJS...74..501P}
{Pinsonneault}, M.~H., {Kawaler}, S.~D., \& {Demarque}, P. 1990, \apjs, 74, 501

\bibitem[{{Prantzos}(2012)}]{Prantzos2012b}
{Prantzos}, N. 2012, \aap, 542, A67

\bibitem[{{Richard} {et~al.}(2005){Richard}, {Michaud}, \&
  {Richer}}]{2005ApJ...619..538R}
{Richard}, O., {Michaud}, G., \& {Richer}, J. 2005, The Astrophysical Journal,
  619, 538

\bibitem[{{Richard} {et~al.}(1996){Richard}, {Vauclair}, {Charbonnel}, \&
  {Dziembowski}}]{1996A&A...312.1000R}
{Richard}, O., {Vauclair}, S., {Charbonnel}, C., \& {Dziembowski}, W.~A. 1996,
  \aap, 312, 1000

\bibitem[{{Richer} {et~al.}(2000){Richer}, {Michaud}, \&
  {Turcotte}}]{2000ApJ...529..338R}
{Richer}, J., {Michaud}, G., \& {Turcotte}, S. 2000, The Astrophysical Journal,
  529, 338

\bibitem[{{R{\"u}diger} {et~al.}(2018){R{\"u}diger}, {Gellert}, {Hollerbach},
  {Schultz}, \& {Stefani}}]{2018PhR...741....1R}
{R{\"u}diger}, G., {Gellert}, M., {Hollerbach}, R., {Schultz}, M., \&
  {Stefani}, F. 2018, \physrep, 741, 1

\bibitem[{{R{\"u}diger} {et~al.}(2015){R{\"u}diger}, {Gellert}, {Spada}, \&
  {Tereshin}}]{2015A&A...573A..80R}
{R{\"u}diger}, G., {Gellert}, M., {Spada}, F., \& {Tereshin}, I. 2015, \aap,
  573, A80

\bibitem[{{R{\"u}diger} {et~al.}(2007){R{\"u}diger}, {Hollerbach}, {Schultz},
  \& {Elstner}}]{2007MNRAS.377.1481R}
{R{\"u}diger}, G., {Hollerbach}, R., {Schultz}, M., \& {Elstner}, D. 2007,
  \mnras, 377, 1481

\bibitem[{{Santos} {et~al.}(2004){Santos}, {Israelian}, {Randich}, {Garc{\'\i}a
  L{\'o}pez}, \& {Rebolo}}]{2004A&A...425.1013S}
{Santos}, N.~C., {Israelian}, G., {Randich}, S., {Garc{\'\i}a L{\'o}pez},
  R.~J., \& {Rebolo}, R. 2004, \aap, 425, 1013

\bibitem[{{Schatzman}(1993)}]{1993A&A...279..431S}
{Schatzman}, E. 1993, \aap, 279, 431

\bibitem[{{Semenova} {et~al.}(2020){Semenova}, {Bergemann}, {Deal},
  {Serenelli}, {Hansen}, {Gallagher}, {Bayo}, {Bensby}, {Bragaglia}, {Carraro},
  {Morbidelli}, {Pancino}, \& {Smiljanic}}]{2020A&A...643A.164S}
{Semenova}, E., {Bergemann}, M., {Deal}, M., {et~al.} 2020, \aap, 643, A164

\bibitem[{{Sestito} \& {Randich}(2005)}]{2005A&A...442..615S}
{Sestito}, P. \& {Randich}, S. 2005, \aap, 442, 615

\bibitem[{{Siess} {et~al.}(2000){Siess}, {Dufour}, \&
  {Forestini}}]{2000A&A...358..593S}
{Siess}, L., {Dufour}, E., \& {Forestini}, M. 2000, \aap, 358, 593

\bibitem[{{Soderblom} {et~al.}(1993){Soderblom}, {Fedele}, {Jones}, {Stauffer},
  \& {Prosser}}]{1993AJ....106.1080S}
{Soderblom}, D.~R., {Fedele}, S.~B., {Jones}, B.~F., {Stauffer}, J.~R., \&
  {Prosser}, C.~F. 1993, \aj, 106, 1080

\bibitem[{{Somers} \& {Pinsonneault}(2016)}]{2016ApJ...829...32S}
{Somers}, G. \& {Pinsonneault}, M.~H. 2016, \apj, 829, 32

\bibitem[{{Spada} {et~al.}(2016){Spada}, {Gellert}, {Arlt}, \&
  {Deheuvels}}]{2016A&A...589A..23S}
{Spada}, F., {Gellert}, M., {Arlt}, R., \& {Deheuvels}, S. 2016, \aap, 589, A23

\bibitem[{{Spite} \& {Spite}(1982)}]{1982A&A...115..357S}
{Spite}, F. \& {Spite}, M. 1982, Astronomy and Astrophysics, 115, 357

\bibitem[{{Spruit}(2002)}]{2002A&A...381..923S}
{Spruit}, H.~C. 2002, \aap, 381, 923

\bibitem[{{Talon} \& {Charbonnel}(2008)}]{2008A&A...482..597T}
{Talon}, S. \& {Charbonnel}, C. 2008, \aap, 482, 597

\bibitem[{{Thoul} {et~al.}(1994){Thoul}, {Bahcall}, \&
  {Loeb}}]{1994ApJ...421..828T}
{Thoul}, A.~A., {Bahcall}, J.~N., \& {Loeb}, A. 1994, \apj, 421, 828

\bibitem[{{Yan} {et~al.}(2021){Yan}, {Zhou}, {Zhang}, {Li}, {Gao}, {Shi},
  {Zhao}, {Aoki}, {Matsuno}, {Li}, {Xu}, {Li}, {Wu}, {Jin}, {Mosser}, {Bi},
  {Fu}, {Pan}, {Suda}, {Liu}, {Zhao}, \& {Liang}}]{2021NatAs...5...86Y}
{Yan}, H.-L., {Zhou}, Y.-T., {Zhang}, X., {et~al.} 2021, Nature Astronomy, 5,
  86

\bibitem[{{Young}(2018)}]{2018ApJ...855...15Y}
{Young}, P.~R. 2018, \apj, 855, 15

\bibitem[{{Zahn}(1992)}]{1992A&A...265..115Z}
{Zahn}, J.~P. 1992, Astronomy and Astrophysics, 265, 115

\end{thebibliography}

\clearpage
\onecolumn

\end{document}